# Structural disorder and magnetic correlations driven by oxygen doping in $Nd_2NiO_{4+\delta}$ ($\delta$ ~ 0.11)

Sumit Ranjan Maity,[1,2] Monica Ceretti,[3] Lukas Keller,[1] Jürg Schefer,[1] Tian Shang,[4,5,6] Ekaterina Pomjakushina,[4] Martin Meven,[7] Denis Sheptyakov,[1] Antonio Cervellino,[5] Werner Paulus[3]

[1]*Laboratory for Neutron Scattering and Imaging, Paul Scherrer Institut, Villigen CH-5232, Switzerland.*

[2]*University of Geneva, Department of Quantum Matter Physics (DQMP) 24, Quai Ernest Ansermet CH-1211 Genève 4, Switzerland*

[3]*Institut Charles Gerhardt Montpellier, UMR 5253, CNRS-UM-ENSCM, Université de Montpellier, 34095 Montpellier, France*

[4]*Laboratory for Multiscale Materials Experiments, Paul Scherrer Institut, Villigen CH-5232, Switzerland*

[5]*Swiss Light Source, Paul Scherrer Institut, Villigen CH-5232, Switzerland*

[6]*Institute of Condensed Matter Physics, École polytechnique fédérale de Lausanne (EPFL), Lausanne CH-1015, Switzerland*

[7]*Institute of Crystallography, RWTH Aachen University and Jülich Centre for Neutron Science (JCNS), Forschungszentrum Jülich GmbH at Heinz Maier-Leibnitz Zentrum (MLZ), 85747 Garching, Germany*

We investigated the influence of oxygen over-stoichiometry on apical oxygen disorder and magnetic correlations in $Nd_2NiO_{4+\delta}$ ($\delta$ ~ 0.11) in the temperature range of 2-300 K by means of synchrotron x-ray powder diffraction, neutron single crystal and powder diffraction studies, combined with macroscopic magnetic measurements. In the investigated temperature range, the compound crystalizes in a tetragonal commensurate structure with the $P4_2/ncm$ space group with excess oxygen atoms occupy the 4*b* (¾ ¼ ¼) interstitial sites, coordinated by four apical oxygen atoms. Large and anisotropic thermal displacement parameters are found for equatorial and apical oxygen atoms, which are strongly reduced on an absolute scale compared to the $Nd_2NiO_{4.23}$ phase. Maximum Entropy analysis of the neutron single crystal diffraction data uncovered anharmonic contributions to the displacement parameters of the apical oxygen atoms, toward the nearest vacant 4*b* interstitial site, related to the phonon assisted oxygen diffusion mechanism. Macroscopic magnetization measurements and neutron powder diffraction studies reveal long-range antiferromagnetic ordering of the Ni-sublattice at $T_N$ ~ 53 K with a weak ferromagnetic component along the *c*-axis, while the long-range magnetic ordering of the Nd-sublattice occurs below 10 K. Temperature dependent neutron diffraction patterns show the appearance of a commensurate magnetic order at $T_N$ with the propagation vector $\boldsymbol{k}$ = (100) and the emergence of an additional incommensurate phase below 30 K, while both phases coexist at 2 K. The commensurate magnetic structure is best described by the $P4_2/nc'm'$ Shubnikov space group. Refined magnetic moments of the Ni and Nd-sites at 2 K are 1.144(76) $\mu_B$ and 1.632(52) $\mu_B$ respectively. A possible origin of the incommensurate phase is discussed and a tentative magnetic phase diagram is proposed.

# I. INTRODUCTION

Perovskite-type layered rare earth Ruddlesden-Popper nickelates with the chemical formula $Ln_2NiO_{4+\delta}$ (Ln: La, Pr, Nd) exhibit high oxygen mobility down to moderate temperatures [1–5]. Therefore, they are promising materials for technological devices such as solid oxide fuel cells, oxygen membranes, sensors, etc. [6–9]. They are also a special class of materials into which oxygen ions can be intercalated reversibly even at ambient temperature using electrochemical methods [10, 11] and the oxygen over-stoichiometry ($\delta$) can be tuned between $0 \leq \delta \leq 0.25$. The crystal structure of these compounds consists of $NiO_2$ layers, which are interspersed between $Ln_2O_2$ rock salt type layers along the long axis (*c*-axis) of the unit cell (Fig. 1). Intercalated extra oxygen atoms are located inside the rock-salt layer in a tetrahedral site, coordinated by four apical oxygen atoms ($O_{ap}$) of two adjacent $NiO_6$ octahedral layers. In these materials, oxygen diffusion proceeds at already moderate temperatures via the apical and vacant interstitial oxygen sites, triggered by low energy phonon modes [12–15]. In this context, large and anharmonic thermal displacement factors were found for apical oxygen atoms of $Pr_2NiO_{4+\delta}$ and its Nd homologue towards the [110]-direction at room temperature, with respect to the average tetragonal F4/*mmm* unit cell [16]. These displacements of the apical oxygen atoms are induced by the presence of interstitial oxygen atoms, creating specific low energy phonon modes, triggering oxygen diffusion down to ambient temperature [12,13,17]. The presence of interstitial oxygen atoms is important, as it yields a symmetric increase of the $O_{int}(O_{ap})_4$ tetrahedra, shifting the $O_{ap}$ atoms towards adjacent vacant interstitial sites, thus creating a shallow energy diffusion pathway between apical and interstitial oxygen sites [16]. This highly anisotropic oxygen diffusion process is essentially different from the thermally activated classical hopping process with more isotropic character, for which elevated energy barriers for oxygen diffusion are difficult to overcome at ambient temperature [15].

Our present work focuses on the $Nd_2NiO_{4+\delta}$ compound, with the aim to better understand the underlying oxygen diffusion mechanism as a function of $\delta$, but also to correlate to changes of the physical properties, i.e. changes in the magnetic properties with oxygen and thus hole doping concentration. The structural phase diagram of $Nd_2NiO_{4+\delta}$ as a function of $\delta$ is complex, as previously revealed by x-ray and neutron diffraction studies [10,18–26]. Stoichiometric $Nd_2NiO_{4.0}$ adopts at 300 K the low-temperature orthorhombic (LTO) phase, with *Bmab* space group. The structural distortion associated with the LTO phase involves systematic canting of $NiO_6$ octahedron around the crystallographic *b*-axis axis. A first order structural phase transition occurs on cooling at 130 K from the LTO phase towards a low temperature tetragonal (LTT) phase with $P4_2/ncm$ space group [25]. The tilting axis of the $NiO_6$ octahedra changes from the [100]-towards the [110]-direction. Contradictory observations are, however, reported regarding the crystal structure of moderately oxygen doped phases [10,18,19,22]. A biphasic model (F*mmm* + P*ccn*) was proposed for $0.1 < \delta < 0.15$ while an orthorhombic P*ccn* phase is reported for $0.067 \leq \delta \leq 0.10$ [18]. In another study, a two phase structural model (B*mab*+I4/*mmm*) for $\delta < 0.07$ and a single tetragonal I4/*mmm* phase with $\delta = 0.07$ was reported [21]. Just recently, a single phase tetragonal $P4_2/ncm$ crystal structure is proposed for the homologous $Pr_2NiO_{4+\delta}$ phase at room-temperature for $0.07 \leq \delta \leq 0.1$ based on in-situ neutron powder diffraction measurements during electrochemical oxidation/reduction [10].

In this work we focus on the LTT phase of $Nd_2NiO_{4+\delta}$ with $\delta \sim 0.11$, rendering formally ¼ of the Ni-atoms in a three-valent oxidation state. This is not only interesting in terms of related electronic properties, but also with respect to the tetragonal symmetry, which is usually observed for much lower $\delta$ only. Structure analysis is carried out combining high-resolution x-ray powder diffraction studies with high-resolution neutron powder diffraction in the temperature range of 2-300 K. In addition, we report on high-resolution single crystal neutron diffraction studies to explore the influence of additional oxygen atoms on apical oxygen disorder scenario in $Nd_2NiO_{4.11}$ at ambient and low temperatures.

Undoped $Nd_2NiO_4$ shows G-type antiferromagnetic order of $Ni^{2+}$ below $T_N = 320$ K with a propagation vector $\boldsymbol{k} = (100)$, with oriented spins parallel to the *a*-axis [25]. In the LTT phase below 130 K, the spin direction of $Ni^{2+}$ changes toward the [110]-direction with a weak ferromagnetic component along the *c*-axis. Furthermore, polarization of the $Nd^{3+}$ ions occurs at temperatures as high as 70 K while cooperative long-range antiferromagnetic ordering of the $Nd^{3+}$ ions takes place below 8 K. Incorporation of additional oxygen atoms frustrates antiferromagnetic superexchange interaction and destroys the long-range commensurate magnetic order at larger doping concentration. Consequently, the oxidized $Nd_2NiO_{4.23}$ phase does not possess any static magnetic order down to 1.5 K [22, 24]. Complex spin ordering schemes are reported for oxygen doped $La_2NiO_{4+\delta}$ phases [28–31] in which the magnetic order changes from commensurate Néel type to incommensurate stripe order at $\delta \sim 0.116$ [32]. However, a clear understanding on the subtle changes in magnetic interactions induced by small changes of $\delta$ is still lacking for the $Ln_2NiO_{4+\delta}$ phases (Ln = Pr, Nd). In addition, anomalous suppression of superconductivity and evidence for static stripe order associated with the unique lattice distortion of LTT structure has been discovered in a crystal of $La_{1.6-x}Nd_{0.4}Sr_xCuO_4$ with $x = 0.12$ [33]. Therefore, in this work, we explore the magnetic properties of the moderately oxygen doped LTT phase of $Nd_2NiO_{4+\delta}$ with $\delta \sim 0.11$ by means of neutron powder diffraction measurements coupled with macroscopic magnetic studies.

## II. EXPERIMENTAL METHODS

Polycrystalline samples were prepared by the classical solid-state reaction route at high temperature starting from stoichiometric $Nd_2O_3$ and NiO powders (99.9% purity, Alfa Aesar). $Nd_2O_3$ was annealed in air at 1173 K overnight to remove any hydroxide traces prior to the synthesis. Mixed starting materials were finely ground, and calcined for 12 h at 1523 K in air. This step was repeated twice with intermediate grinding and pelletizing. When prepared under air at high temperature, the overall $\delta$ value in the as-grown powder sample is ~ 0.22-0.25 [30]. Therefore, the final powder sample with reduced oxygen non-stoichiometry was achieved by reduction of the as-prepared sample in vacuum ($10^{-3}$ mbar) at 1073 K for 48 h.

The overall oxygen content was determined by thermogravimetric analysis in 5%-$H_2$/95%-He gas atmosphere in the temperature range of 300-1250 K (see Fig. S-1 in the Supplemental Material at [35]). The measurement was carried out on a NETZSCH STA 449C analyzer equipped with PFEIFFER VACUUM ThermoStar mass spectrometer. The sample was fully decomposed into $Nd_2O_3$ and metallic Ni at high temperature. From the weight loss, the average $\delta$ value in the reduced powder sample was determined to be 0.11(2). In the following, we designated the corresponding sample as $Nd_2NiO_{4.11}$.

Synchrotron x-ray powder diffraction (SXRPD) measurements were performed at different temperatures in between 5 and 300 K at the Material Sciences (MS) beamline X04SA at the Swiss light source of the Paul Scherrer Institute (PSI), Switzerland [36]. Measurements were performed with a nominal photon energy of 17.5 keV. The exact wavelength ($\lambda$ = 0.6208(1) Å) and instrumental resolution parameters were determined from a standard $LaB_6$ powder (NIST), measured under identical experimental conditions. Data were collected in the range of $3^\circ \leq 2\theta \leq 120^\circ$ with a step size of 0.0036˚. Samples were filled into a thin glass capillary of 0.3 mm diameter and cooled with liquid He in a Janis Cryostat. During data collection, the sample was continuously rotated to reduce the effect of preferred orientation. Neutron powder diffraction (NPD) patterns were collected at variable temperatures on the thermal neutron diffractometer HRPT ($\lambda$ = 1.4940(2) Å) [37] and on the cold neutron diffractometer DMC ($\lambda$ = 2.4586(3) Å) [38,39] at PSI. The powder sample was mounted using a thin-walled vanadium can of diameter 8 mm. The sample was cooled in a top-loading $^{3}$He cryostat achieving temperatures down to 1.7 K. All powder patterns were analyzed by Rietveld method using the *FullProf suite* program [40].

The dc magnetic susceptibility (both zero-field-cooled and field-cooled) measurements were performed as a function of temperature during the slow heating of the sample from 2-320 K with an applied magnetic field of 1 Tesla using the Magnetic Property Measurement System (MPMS) at Laboratory for Scientific Developments and Novel Materials (LDM), PSI. Isothermal magnetization measurements were performed as a function of applied magnetic field with decreasing fields from 9 T using the Physical Property Measurement System (PPMS) at LDM, PSI.

The single crystal sample was grown from the corresponding powder rods as described elsewhere [34]. A small section (~ 200 mg) was cut and separately reduced in vacuum as previously described for the powder sample. Single crystal neutron diffraction measurements were performed on the hot-neutron four-circle diffractometer HEiDi (equipped with a point detector) at the Heinz Maier-Leibnitz Zentrum (MLZ) at the FRM-II reactor in Garching, Germany [41]. A short wavelength ($\lambda$ = 0.793(1) Å) was used for data collection up to high momentum transfers ($q = \sin\theta/\lambda \sim 0.9$ Å$^{-1}$). P-type reflections of (–*hkl*) and (*hkl*) type were measured covering two quadrants of the Ewald sphere. Integrated intensities of totally 1291 and 1121 structural Bragg reflections were collected at 300 K and 55 K, respectively. Least-square refinements of the single crystal data were carried out using the *JANA2006* software package [42]. Nuclear densities were reconstructed in real space through the Maximum

Entropy Method (MEM) [43,44] *via* the *DYSNOMIA* program [45] using the limited-memory BFGS optimization algorithm [46]. Crystal structure and nuclear density maps were visualized with the *VESTA* software suite [41, 42].

## III. RESULTS AND DISCUSSION

### A. Crystal structure and oxygen disorder

The room-temperature SXRPD pattern of $Nd_2NiO_{4.11}$ is shown in Fig. 2(a) along with the Rietveld refinement profile in the tetragonal LTT-type structure model. The powder pattern is composed of sharp structural Bragg peaks indicating good crystalline quality of the sample. The refinement reflects the single-phase nature of the sample without any detectable impurity. The lattice parameters are found to be $a = b = 5.46200(2)$ Å, $c = 12.20517(9)$ Å at 300 K. No splitting of (*h0l*)/(*0kl*) type Bragg reflections was detected down to 5 K (see Fig. S-2 in the Supplemental Material at [35]) within the high-resolution of our synchrotron data, indicating the LTT-type crystal structure in the entire temperature range of 2-300 K. In the LTT model, Nd, Ni, $O_{ap}$, $O_{eq1}$, $O_{eq2}$, and $O_{int}$ atoms occupy 8*i* (*x x z*), 4*d* (0 0 0), 8*i* (*x x z*), 4*a* (3/4 1/4 0), 4*e* (1/4 1/4 *z*), and 4*b* (3/4 1/4 1/4) Wyckoff sites, respectively. The refined room temperature crystal structure is visualized in Fig. 1. Structural and thermal parameters obtained with Rietveld refinements are summarized in table S-1 and corresponding NPD patterns along with Rietveld refinement profiles are presented in Fig. S-3 of the Supplemental Material at [35]. Note that, on average only 11% of all 4*b* sites are occupied by interstitial oxygen atoms in $Nd_2NiO_{4.11}$, approximately corresponding to ½ excess oxygen atom per unit cell. The LTT-type crystal structure is characterized by a systematic tilting of the $NiO_6$ octahedron around the [110]-direction. The tilt angle $\alpha_1$ is defined by the angle between the *c*-axis and the respective shift of the $O_{ap}$, as presented in Fig. 1(b). Extracted values of $\alpha_1$ at 300 K and 2 K are 9.3˚ and 9.6˚, respectively. For the corresponding LTT structure of stoichiometric $Nd_2NiO_{4.0}$ [25,49], the respective tilt angle was found to be 10.3˚ at 2 K. The LTT-type tilt pattern generates two different sets of equatorial oxygen atoms, $O_{eq1}$ atoms at basal plane while $O_{eq2}$ atoms are slightly displaced along the *c*-axis. Such a displacement of $O_{eq2}$ atoms, further characterized by the angle $\alpha_2$, lifts the center of inversion between the two in plane magnetic $Ni^{2+}$ ions that consequently induces an antisymmetric Dzyaloshinskii-Moriya (DM) interaction. The spin canting resulting from the DM interactions gives rise to the weak ferromagnetic properties in the LTT structure of $Nd_2NiO_{4.11}$ below the magnetic ordering temperature as discussed with macroscopic magnetization measurements in the next section. Refined values of $\alpha_2$ at 300 K and 2 K are 7.7˚ and 8.4˚ while for stoichiometric $Nd_2NiO_{4.0}$, it is found to be 10.5˚ at 2 K. It is at first sight surprising to find the LTT-type structure for $Nd_2NiO_{4.11}$ compound, since no significant disorder is introduced in the long-range $NiO_6$ octahedron tilting pattern by such a considerable amount of interstitial oxygens. Furthermore, no superstructure reflections were observed in the x-ray powder diffraction patterns down to 2 K, indicating the absence of long-range ordering of excess oxygen atoms in the investigated temperature range. This observation is in strong contrast with that reported for iso-structural $La_2NiO_{4+\delta}$ in which one-dimensional (1D) ordering of excess oxygens is observed for similar concentrations [33].

The temperature dependence of the unit cell volume (inset of Fig. 2(a)) is fitted with a quasi-harmonic Debye model [50] (Eq. (1))

$$V(T) = V_0 + A \cdot \left(\frac{T}{\Theta_D}\right)^3 \cdot T \int_0^{\Theta_D/T} \frac{x^3}{e^x - 1}\, dx \quad (1)$$

where $A$ and $V_0$ are constants and $\Theta_D$ is the Debye temperature, indicating the thermal expansion to be mainly dominated by phonon contributions. The Debye temperature could be fitted to $\Theta_D$ = 414(41) K in agreement with that reported for other $K_2NiF_4$-type nickelates [44].

For a more detailed structural analysis, especially in view of a precise analysis of the apical oxygen displacements, least-square refinements of single crystal neutron diffraction were performed. Results are summarized in table I and are in good agreement with those obtained from the powder diffraction data. The $\delta$ value in the single crystal is refined to be 0.097(10). A detailed inspection of the apical oxygen atoms yield an occupancy corresponding to an overall apical site stoichiometry of 1.80(1) at ambient and 1.71(2) at 55K, i.e. 10% and 15% less than the stoichiometric occupation, while at the same time the occupancies for the equatorial oxygen atoms correspond very closely to the ideal stoichiometry. The presumed under-stoichiometry is, however, not real but an artefact, related to strong and anharmonic distortions of the $O_{ap}$, which can no longer be described in a harmonic approach.

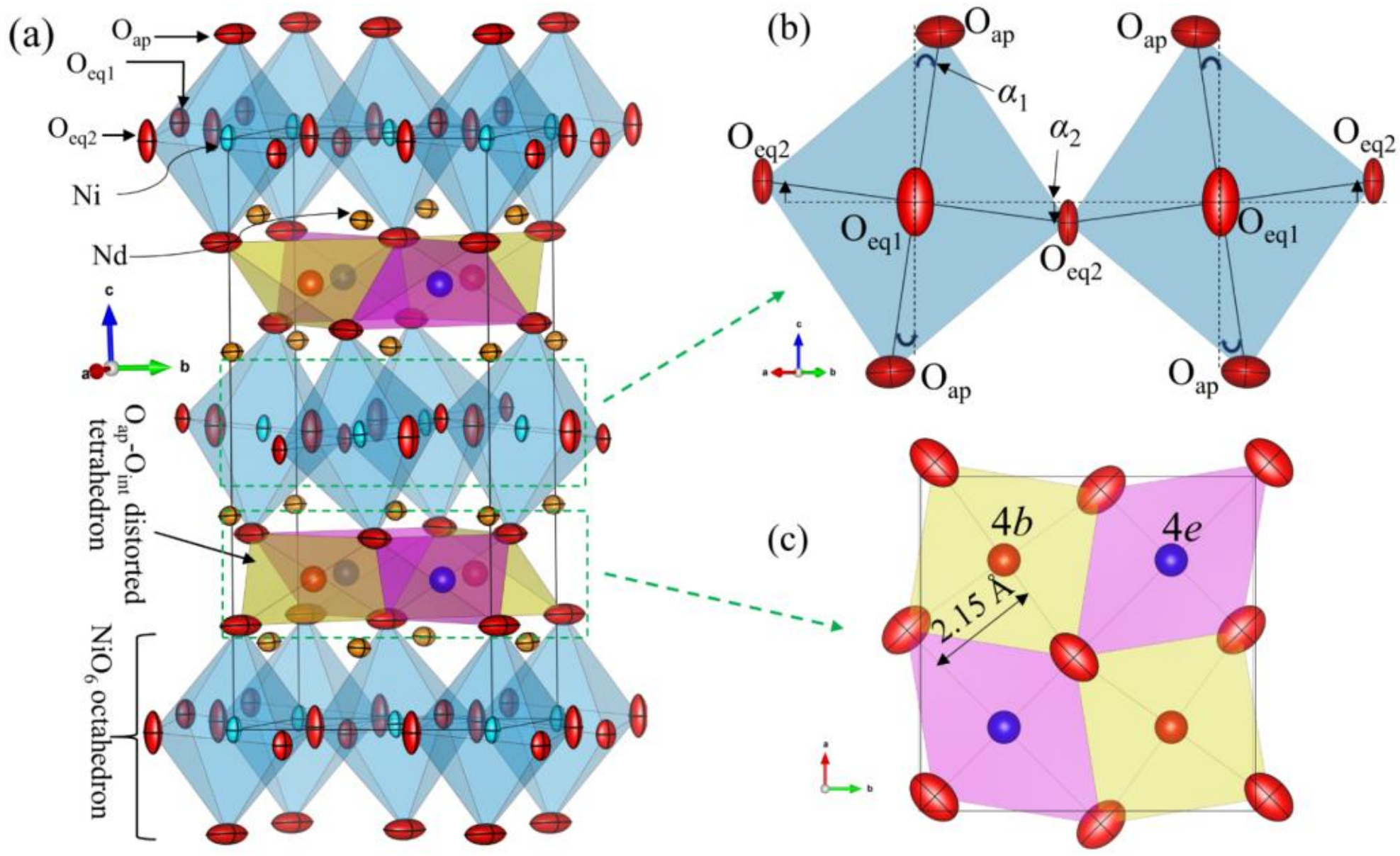

**Figure 1:** (a) Perspective view of the tetragonal unit cell of $Nd_2NiO_{4.11}$ in $P4_2/ncm$ space group, showing two distinct tetrahedral sites 4*b* (3/4, 1/4, 1/4) (red spheres) and 4*e* (1/4, 1/4, *z* ~ 1/4)

(blue spheres) accessible for extra oxygen atoms, while only the 4*b* Wyckoff site with point symmetry $\bar{4}$ is occupied and the 4*e* tetrahedral site with point symmetry 2.*mm* being entirely empty. Note that, on average only ½ excess oxygen atom is present per unit cell i.e. only 11% of 4*b* sites are filled in $Nd_2NiO_{4.11}$. (b) [110]/[001] projection of $NiO_6$ octahedral layer at $z$ = 0.5 showing specifically the $NiO_6$ octahedron tilt pattern about the [110]-axis. (c) [100]/[010] projection of the two $O_{int}(O_{ap})_4$ tetrahedrons with Wyckoff positions 4*e* and 4*b* showing in- and exhaling behaviour of the empty (pink) and occupied (yellow) tetrahedra, respectively.

The crystal structure using harmonic displacement factors (probability 85%) is visualized in Fig. 1, while a more sophisticated data analysis using the Maximum Entropy Method is given in Figs. 3(a)-(e), also comparing to the NPD data. The refinement given in Table I is used for phasing the respective intensities for the Maximum Entropy analysis. It thus becomes clear that all apical oxygen atoms are strongly displaced from their equilibrium positions in a way to allow a symmetrical expansion of the $O_{int}(O_{ap})_4$ tetrahedra, yielding an increased $O_{int}$-$O_{ap}$ distance of 2.76 Å at 55K, compared to only 2.15 Å when using the average structural model as presented in Fig.1. The as obtained real $O_{int}$-$O_{ap}$ distance of 2.76Å corresponds very well to the expected O-O distance, taking into account an $O^{2-}$ ionic radius of 1.4 Å according to Shannon [52]. These findings also indicate strong displacements of $O_{ap}$ in the [110]-direction inside the rock salt layer, favoring an easy oxygen diffusion pathway between apical and vacant interstitial sites, in agreement with a theoretically predicted push-pull diffusion mechanism reported in [14,53]. From a methodological aspect, it is also interesting to state that only high-resolution diffraction data up to 0.9 $Å^{-1}$ allowed to visualize such important structural details, which do not emerge from the NPD data, obtained up to 0.66 $Å^{-1}$ only. It is noteworthy that a similar disorder scenario of the apical oxygen atoms has been reported for $Pr_2NiO_{4.09}$ at room-temperature [10], from single crystal neutron diffraction studies, suggesting this type of disorder to be a general feature, commonly occurring in $K_2NiF_4$-type oxides [54,55]. However, related $O_{ap}$ displacements become much more obvious in $Nd_2NiO_{4.11}$, probably related to the lower ionic radius of $Nd^{3+}$ (1.163 Å compared to 1.179 Å for $Pr^{3+}$ in IX-fold coordination). The same argument can also be taken to explain the different transition temperatures for the orthorhombic to tetragonal phase transition comparing $Pr_2NiO_{4.23}$ and $Nd_2NiO_{4.23}$, being 638K and 1003K respectively. Fig. 3(f) represents a 1D cut through the nuclear density maps of apical oxygen atoms, showing precisely the temperature dependence of the apical oxygen disorder. Two localized nuclear density lobes, separated by a distance of 0.621 Å from the apical oxygen center along [110]-direction, are clearly evident at 55 K. These two nuclear density lobes behave like splitted apical sites at 55 K. However, with increasing temperature, a clear overlap of nuclear densities of these two lobes with the center occurs suggesting the movements related to the libration-mode of $NiO_6$-octahedra and an onset of phonon assisted oxygen diffusion process at 300 K.

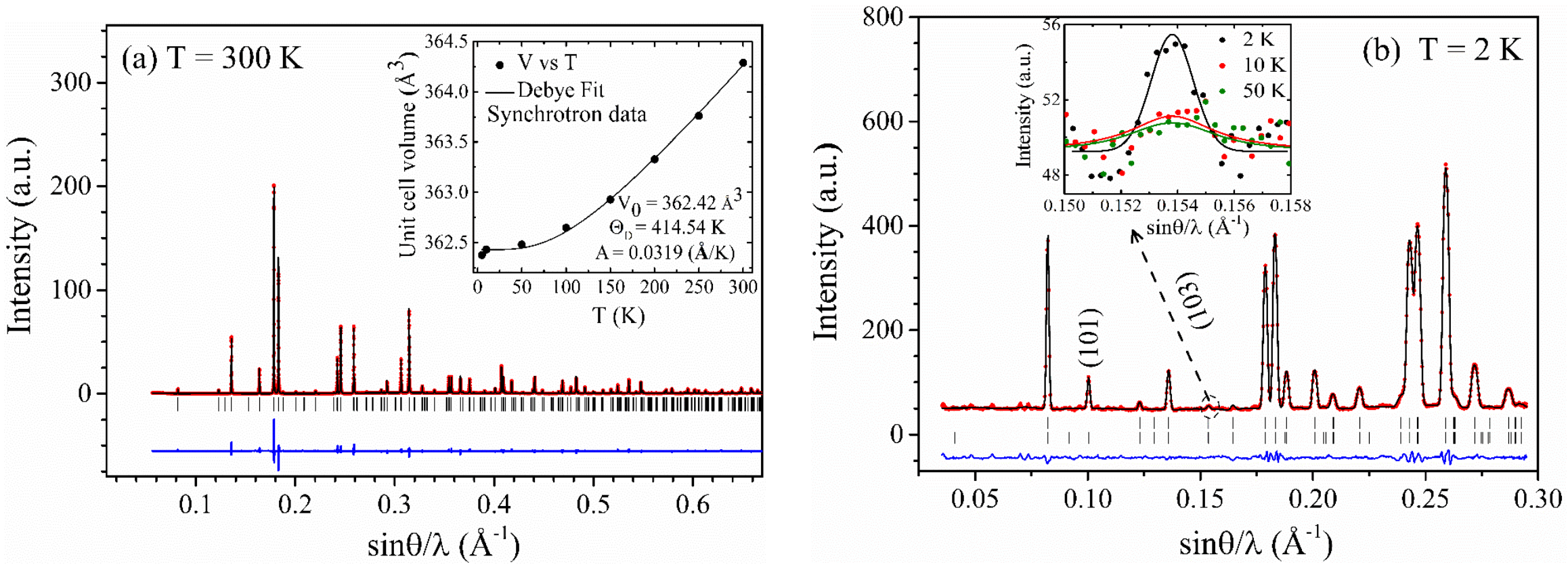

**Figure 2:** Observed (red circles), calculated (black line) and difference (blue continuous line) patterns resulting from Rietveld analysis of (a) SXRPD data collected at 300 K and (b) neutron diffraction data collected on DMC at 2 K for $Nd_2NiO_{4.11}$. For the low temperature neutron data, upper and lower row of ticks indicate the positions of structural and magnetic Bragg reflections, respectively. The difference curve is on the same scale but shifted down for clear visualization. The inset of panel (a) shows the temperature evolution of the unit cell volume determined with SXRPD data. Filled circles are experimental data points and the solid line is a fit with the Debye model (Eq. (1)). The inset of panel (b) shows the intensity of the (103) reflection showing only magnetic contribution, indicating the magnetic ordering of Nd-site between 2 and 10 K. Solid lines are only a guide to the eye. The position of strong (101) magnetic peak, essentially related to magnetic intensities for the Ni atoms, is also marked.

**TABLE I:** Least-square refinement results of single crystal neutron diffraction data obtained for the $Nd_2NiO_{4.11}$ compound on HEiDi at MLZ/FRM II, using a wavelength of 0.793(1) Å up to $\sin\theta_{max}/\lambda$ = 0.9 Å$^{-1}$. Refinements were carried out in $P4_2/ncm$ space group. Displacement parameters $U_{ij}$ are given in Å$^2$.

| Temperature | | 300 K | 55 K |
|---|---|---|---|
| $a = b$ | (Å) | 5.4597(5) | 5.4498(2) |
| $c$ | (Å) | 12.2027(4) | 12.1536(6) |
| Ni (4*d*) (0 0 0) | Occ. | 1 | 1 |
| | $U_{11} = U_{22}$ | 0.0045(3) | 0.0032(5) |
| | $U_{33}$ | 0.0132(5) | 0.0073(6) |
| | $U_{12}$ | -0.0006(2) | -0.0005(2) |
| | $U_{13} = U_{23}$ | 0.00049(18) | 0.0003(2) |
| Nd (8*i*) (*x x z*) | Occ. | 2 | 2 |
| | $x$ | 0.98967(15), | 0.98904(19), |
| | $z$ | 0.36133(7) | 0.36146(8) |
| | $U_{11} = U_{22}$ | 0.0105(4) | 0.0070(5) |
| | $U_{33}$ | 0.0064(4) | 0.0027(6) |
| | $U_{12}$ | 0.0015(2) | 0.0011(3) |
| | $U_{13} = U_{23}$ | 0.00035(17) | 0.0007(2) |
| $O_{ap}$ (8*i*) (*x x z*) | Occ. | 1.800(15) | 1.710(18) |
| | $x$ | 0.0413(3), | 0.0421(4), |
| | $z$ | 0.17706(12) | 0.17681(18) |
| | $U_{11} = U_{22}$ | 0.0287(7) | 0.0200(9) |
| | $U_{33}$ | 0.0078(7) | 0.0060(9) |
| | $U_{12}$ | -0.0097(8) | -0.0071(9) |
| | $U_{13} = U_{23}$ | 0.0005(3) | 0.0000(3) |
| $O_{eq1}$ (4*a*) (¾ ¼ 0) | Occ. | 0.96(1) | 0.95(2) |
| | $U_{11} = U_{22}$ | 0.0060(5) | 0.0054(7) |
| | $U_{33}$ | 0.0325(10) | 0.0270(12) |
| | $U_{12}$ | 0.0020(5) | 0.0001(5) |
| $O_{eq2}$ (4*e*) (¼ ¼ *z*) | Occ | 1.005(15) | 1.04(2) |
| | $z$ | 0.97828(13) | 0.97690(17) |
| | $U_{11} = U_{22}$ | 0.0073(5) | 0.0030(7) |
| | $U_{33}$ | 0.0174(7) | 0.0112(9) |
| | $U_{12}$ | -0.0029(5) | 0.0016(5) |
| $O_{int}$ (4*b*) | Occ. | 0.097(10) | 0.057(14) |
| (¾ ¼ ¼) | $U_{iso}$ | 0.014(3) | 0.008(8) |
| no. of | all | 1290 | 1121 |
| reflections | unique | 670 | 470 |
| $R_p$ (%) | | 5.37 | 6.54 |
| $wR_p$ (%) | | 7.76 | 10.46 |
| $R_{Bragg}$ (%) | | 2.31 | 2.02 |

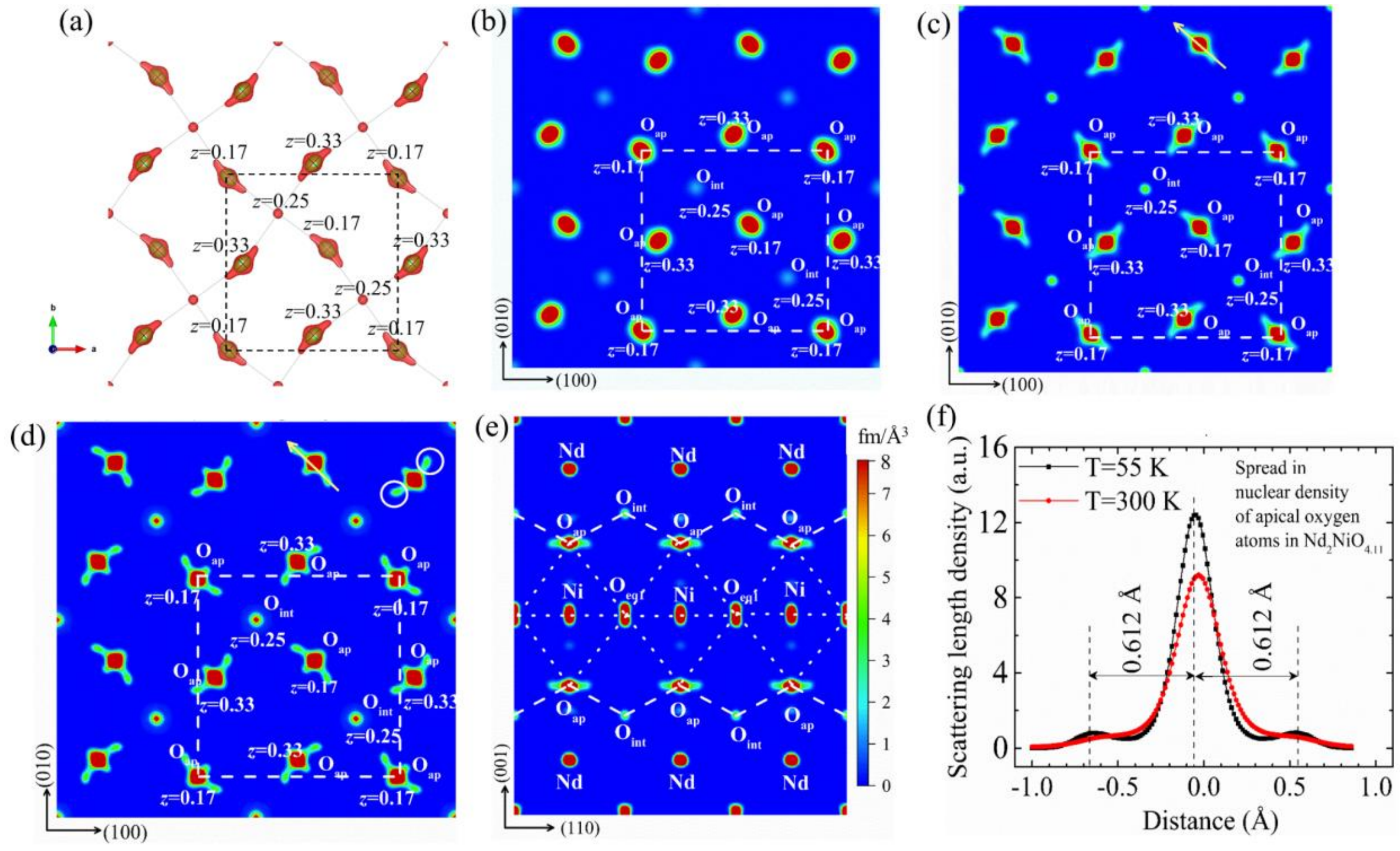

**Figure 3:** MEM reconstructions of $Nd_2NiO_{4.11}$ obtained from single crystal neutron diffraction. (a) Nuclear density iso-surfaces of apical and interstitial oxygen atoms at 300 K. Anisotropic thermal ellipsoids of apical oxygen atoms are also shown. The anharmonic contribution to the shifts of apical oxygen atoms toward nearest 4*b* interstitial oxygen atoms in the (*a*,*b*)-plane is clearly evident. (b) $2a \times 2b$ projection of nuclear density maps obtained by high-resolution neutron powder diffraction data at ambient and (c-d) by high-resolution single crystal neutron diffraction data at 55 K and 300 K, respectively. Cuts were made along the *c*-axis in the range of $0.16 \leq z \leq 0.34$. The tetragonal *P*-centered unit cell is marked with a dashed square. Positions of apical and interstitial oxygen atoms along the *c*-axis are indicated. (e) [110]/[001] projection

cut along the $[1\bar{1}0]$-direction in the range of -0.25 ≤ $z$ ≤ 0.25 showing the oxygen diffusion $O_{ap}$-$O_{int}$-$O_{ap}$ pathway (white dashes) along the [110]-direction obtained with single crystal neutron diffraction data at 300 K. $NiO_6$-octahedra are indicated by white dotted lines. All color maps are plotted on the same scale as shown in (e). (f) Scattering densities of the apical oxygen atoms at different temperatures. Line cuts were made at 55 K and 300 K data through the apical oxygen atom along the [110]-direction, the partial split positions of $O_{ap}$ atoms are clearly evident at 55K.

## B. Magnetic properties

### *1. Macroscopic magnetization*

DC magnetic susceptibility ($\chi = M / H$) curves were recorded both in zero-field-cooled (ZFC) and field cooled (FC) configurations in the temperature range of 2-320 K with 1 T magnetic field. Inverse magnetic susceptibility curves are presented in Fig. 4(a). An anomaly is observed in both curves around the magnetic ordering temperature $T_N$ related to the long-range antiferromagnetic ordering of the Ni-sublattice. The magnetic ordering temperature $T_N$ is found to be 53 K from the $d\chi^{-1}/dT$ curve as shown in the inset of Fig. 4(a). Thus, additional oxygen doping largely suppresses $T_N$ as it is reduced from 330 K in $Nd_2NiO_4$ [56,57] to 53 K in $Nd_2NiO_{4.11}$. However, we did not find any evidence of long-range magnetic ordering of the Nd-sublattice in the susceptibility data. Above 150 K, a linear fit of $\chi^{-1}$ as a function of temperature was made to the following Curie-Weiss law:

$$\chi(T) = \chi_0 + \frac{C}{T - \Theta} \quad (2)$$

From the least-square fitting, we extract $\mu_{\text{eff}}$ = 5.35 $\mu_B$, which in the complete ionic approximation leads to $\mu_{\text{eff}}$ (Ni-site) =1.46 $\mu_B$ taking $Nd^{3+}$ as a free ion (3.63 $\mu_B$) in this temperature range. The reduced value of the effective magnetic moment for the Ni-site is formally due to the presence of either $Ni^{3+}$ $(3d^7)$ or Ni ions in $3d^8$ configuration that is $\left(3d^8\underline{L}\right)$ where $\underline{L}$ is an O 2*p* hole. The Curie-Weiss temperature of $\Theta$ = -36 K indicates predominant antiferromagnetic exchange interactions at low temperature in $Nd_2NiO_{4.11}$. The temperature independent term, $\chi_0$ is found to be $1.3\times10^{-3}$ emu. $mol^{-1}$. A temperature independent contribution can be due to a combination of Pauli paramagnetism, Van Vleck paramagnetism, and core diamagnetism. As our sample is non-metallic, the Pauli paramagnetic contribution is expected to be very small and the core diamagnetic contribution is estimated to be ~ $-0.1\times10^{-3}$ emu. $mol^{-1}$. Therefore, the temperature independent term is mainly dominated by the Van Vleck contribution resulting from the crystal field levels of $Nd^{3+}$ and from $Ni^{2+}$ ions.

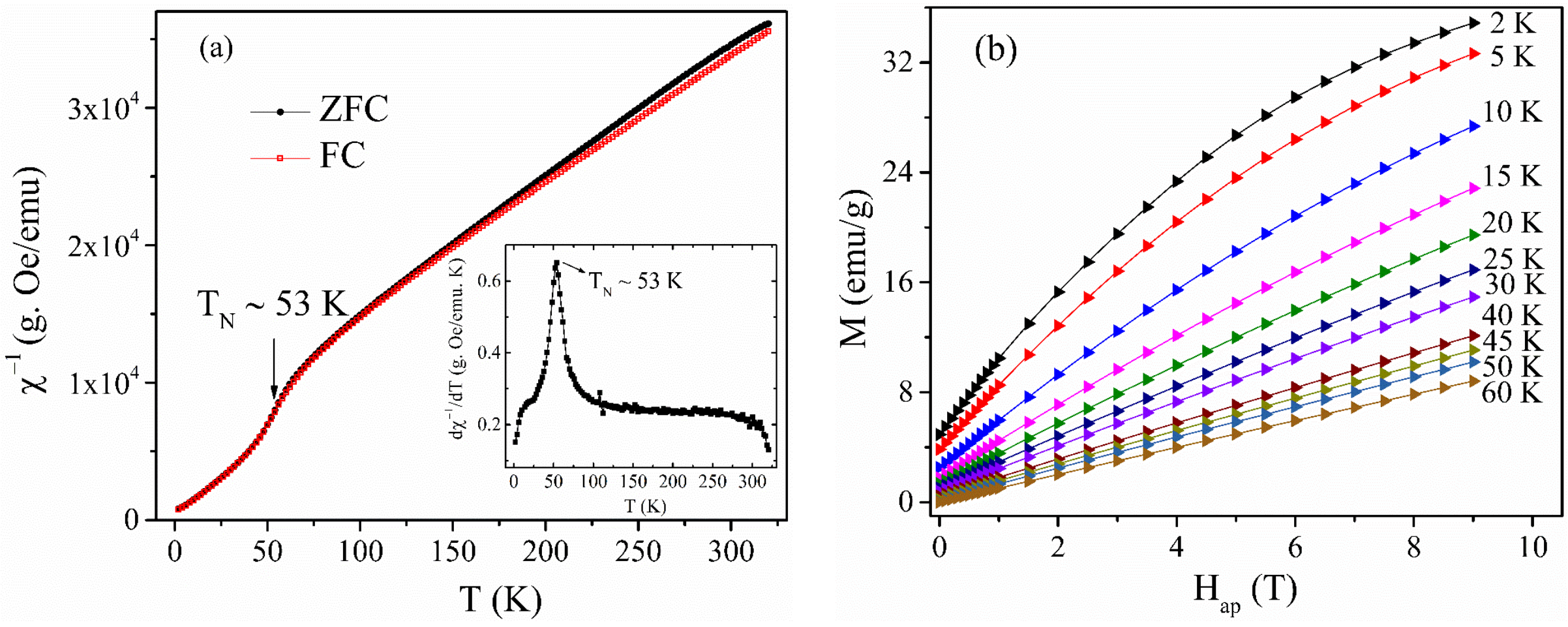

**Figure 4:** (a) Inverse dc magnetic susceptibility curves measured as a function of temperature for the $Nd_2NiO_{4.11}$ powder sample in zero-field cooled (ZFC, black closed circles) and field cooled (FC, red open squares) configurations. The magnetic ordering temperature of the Ni-sublatice is marked with an arrow around 53 K. The inset shows the 1st order temperature derivative of inverse magnetic susceptibility illustrating a sharp magnetic transition around 53

K. (b) Isothermal magnetization curves measured as a function of applied magnetic field between 0-9 Tesla recorded below 60 K for $Nd_2NiO_{4.11}$ powder sample. Note the appearance of a small ferromagnetic component in the magnetization curves below 50 K.

Fig. 4(b) shows several isothermal magnetization curves measured between 2 K and 60 K as a function of magnetic field between 0-9 T. In the high-temperature-range ($T \geq 60$ K), magnetization curves have a linear field dependence with no remanence at all, as expected for the paramagnetic region. However, a small spontaneous magnetization starts to appear around 50 K, showing typical features of a weak ferromagnet. No signature of a magnetic field induced magnetic transition was observed at low temperature as found in the undoped sample [56]. It is worth noting that the magnetization curve shows no indication of magnetic saturation even at 2 K with 9 T magnetic field. However, *M-H* curves become nonlinear below 10 K which possibly indicates a magnetic contribution of $Nd^{3+}$ below this temperature. Magnetic ordering of the Nd-sublattice is further confirmed by NPD data as discussed in the next section. The isothermal magnetization curves (*M-H*) can be represented by the following equation [56]:

$$M(H_a, T) = M_0(T) + \chi_{dif}(T) H_a \quad (3)$$

where $\chi_{dif}(T)$ is the high-field differential susceptibility, $H_a$ the applied magnetic field and $M_0(T)$ the extrapolated spontaneous magnetization. Linear fits were made in the 0-3 T region so that the extrapolated spontaneous magnetization is close to the remanence at each temperature. The temperature dependence of $M_0(T)$ is shown in Fig. 5(a). $M_0(T)$ starts to depart from zero just below 50 K and smoothly increases down to 25 K followed by a strong increase down to 2 K. Similar behavior of $M_0(T)$ in undoped $Nd_2NiO_4$ was described by the evolution of the weak ferromagnetic component of the $Ni^{2+}$ magnetic moments along the *c*-axis [56]. Furthermore, this weak ferromagnetic component polarizes the $Nd^{3+}$ magnetic moments (through the Nd-Ni interactions), which contribute to the observed increase of the spontaneous magnetization. The full ferromagnetic hysteresis curve obtained at 2 K is shown in the inset of Fig. 5(a), where it is evident that the remanence magnetization (~ 4.94 emu/g) and coercive field (~ 0.3 T) are strongly decreased with oxygen doping. As for undoped $Nd_2NiO_4$ [56], the remanence is ~ 15 emu/g and the coercive field is higher than 1 T at 4.5 K. The reduction of the coercive field could be associated with the decrease in magnetocrystalline anisotropy upon additional oxygen doping. The internal field acting on the $Nd^{3+}$ ions due to $Ni^{2+}$ ferromagnetic component is evaluated using the Eq. (4) [56]

$$M^0(T) = M_{Ni}^0 + \chi_{dif}(T)H_i^0 \quad (4)$$

where $H_i^0$ is the internal field induced by the weak ferromagnetic component of $Ni^{2+}$ magnetic moments and $M_{Ni}^0$ is the ferromagnetic component of magnetization from the Ni-sublattice. This relation is depicted in Fig. 5(b), where a linear dependence between $M(T)$ and $\chi_{dif}(T)$ in the temperature range from $T$ = 40 K down to $T$ = 10 K is evident. From the slope and intercept, we get $M_{\mathrm{Ni}}^0$ = -0.08(4) $\mu_B$/f. u. and $H_i^0$ = 1.12(8) T. In the next section, with a group theoretical analysis, we show that the weak ferromagnetic component of magnetization ($M_{Ni}^0$) points towards the crystallographic $c$-axis. Therefore, the internal magnetic field $H_i^0$ acting on $Nd^{3+}$ induced by this ferromagnetic component also points towards the crystallographic $c$-axis. The negative sign of $M_{\mathrm{Ni}}^0$ indicates that the Nd-Ni interaction is antiferromagnetic. A similar scenario is also observed in undoped $Nd_2NiO_{4.0}$ ($M_{\mathrm{Ni}}^0$ = -0.36 $\mu_B$/f. u. and $H_i^0$ = 5.26 T in $Nd_2NiO_{4.0}$) [56], although these values are strongly reduced with oxygen doping. Strong reduction of the ferromagnetic component in the oxygen doped sample can be attributed to the decrease of the DM interactions between in-plane nearest neighbor $Ni^{2+}$ ions. To a first approximation, the spin Hamiltonian can be described by [56]

$$\mathcal{H} = J_{\mathrm{Ni-Ni}} \sum_{\mathrm{NN}} \mathbf{S}_{\mathrm{i}}^{\mathrm{Ni}} . \mathbf{S}_{\mathrm{j}}^{\mathrm{Ni}} + J'_{Ni-Ni} \sum_{\mathrm{NNN}} \mathbf{S}_{\mathrm{i}}^{\mathrm{Ni}} . \mathbf{S}_{\mathrm{j}}^{\mathrm{Ni}} + \boldsymbol{D}_{\mathrm{Ni-Ni}} \sum_{\mathrm{NN}} \mathbf{S}_{\mathrm{i}}^{\mathrm{Ni}} \times \mathbf{S}_{\mathrm{j}}^{\mathrm{Ni}} + J_{\mathrm{Nd-Ni}} \sum_{\mathrm{NN}} \mathbf{S}_{\mathrm{i}}^{\mathrm{Nd}} . \mathbf{S}_{\mathrm{j}}^{\mathrm{Ni}} \qquad \mathbf{(5)}$$

where all the sums are extended to nearest neighbours, except for the second one which is extended to next nearest neighbours. $J_{\text{Ni-Ni}}$ and $D_{\text{Ni-Ni}}$ stands for the in-plane symmetric and antisymmetric DM Ni-Ni superexchange interactions, respectively. $J'_{\text{Ni-Ni}}$ stands for the magnetic interactions between two adjacent $NiO_2$ planes along the *c*-axis, while $J_{\text{Nd-Ni}}$ is the isotropic symmetric antiferromagnetic Nd-Ni superexchange coupling constant (*cf.* Fig. 6(b)). In Eq. (5), the magnetocrystalline anisotropy and Nd-Nd superexchange terms are ignored for simplification. In a first order approximation, the magnitude of the DM interaction is directly proportional to the displacement of $O_{eq2}$ atoms from the basal plane, i.e. largely depends on the angle $\alpha_2$ as shown in Fig. 1(b) and further outlined in Fig. 6(c). Rietveld refinements of NPD data of $Nd_2NiO_{4.11}$ revealed that $\alpha_2$ is decreased by 2˚ at 2 K compared to the $Nd_2NiO_{4.0}$, which is consistent with the observed decrease in the magnitude of the ferromagnetic component. Therefore, our NPD and magnetization data reveals a strong modification in the DM interactions, allowing a spin canting, through the subtle change in the tilting pattern of $NiO_6$ octahedra of LTT structure with oxygen doping in $Ln_2NiO_{4+\delta}$ compounds.

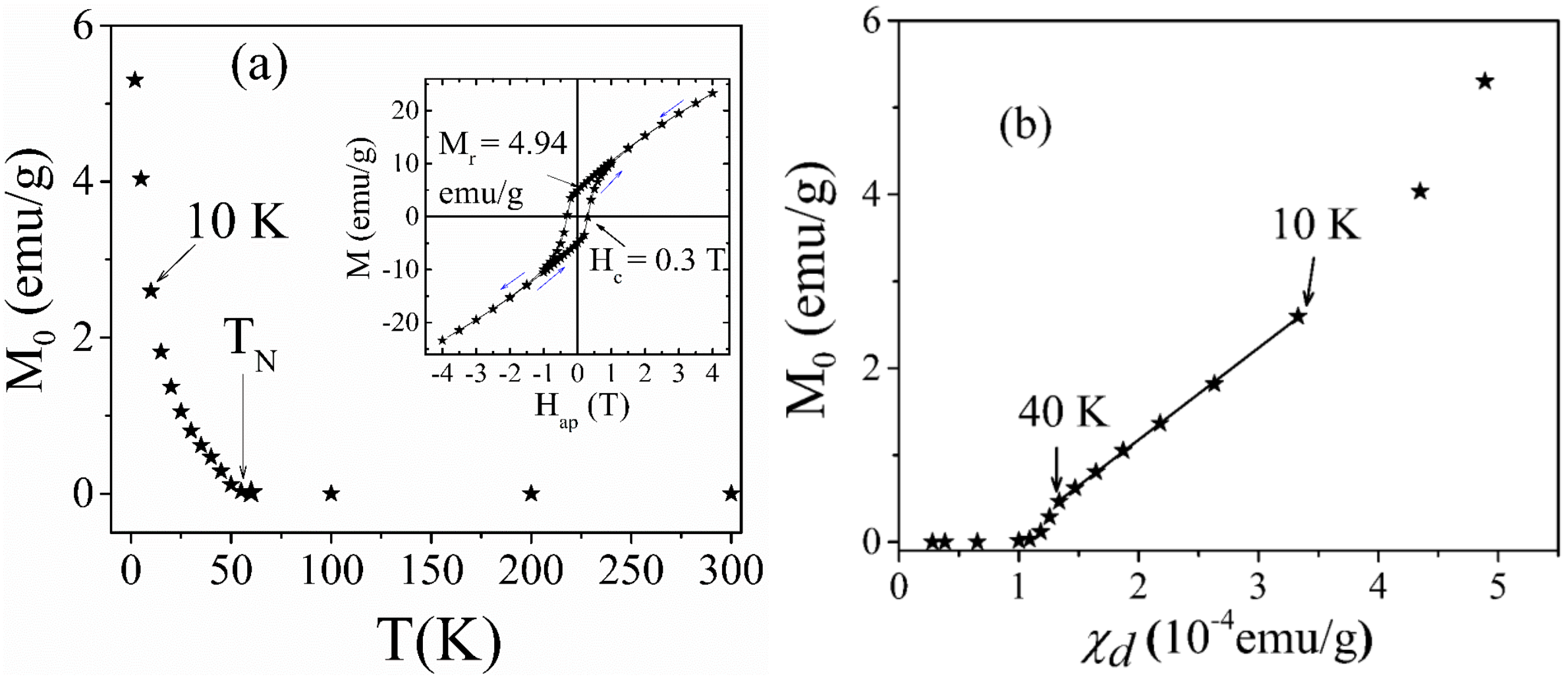

**Figure 5:** (a) Temperature dependence of $M_0$(T), deduced from isothermal magnetization data of $Nd_2NiO_{4.11}$. Néel temperature $T_N$ is marked with an arrow. The inset presents the ferromagnetic hysteresis loop of $Nd_2NiO_{4.11}$, recorded at 2 K. (b) $M_0$(T) *vs* $\chi_d$ curve as discussed in the text. A linear fit is made between 10-40 K to extract the ferromagnetic component $M_{Ni}^0$ and the internal magnetic field $H_i^0$.

### *2. Magnetic structure*

Temperature dependent NPD data reveals the appearance of commensurate superstructure reflections below 50 K. These reflections can be indexed with a propagation vector $\boldsymbol{k}$ = (100) with respect to the tetragonal chemical unit cell. These reflections are related to the long-range magnetic ordering of Ni-sublattice. Fig. 6(a) shows the variation of the integrated intensity of (101) magnetic reflection with temperature. The (101) magnetic reflection is predominantly related to contributions of Ni-magnetic moments. Therefore, the temperature evolution of (101) magnetic reflection confirms the onset of magnetic ordering of Ni-sublattice around 50 K, in agreement with our dc magnetic susceptibility data. The concave upward shape of the integrated intensity of the (101) reflection indicates that the magnetic moment for the Ni is not saturated, even at 2 K. The magnetic ordering of the Nd-sublattice below 10 K is confirmed from the temperature evolution of NPD patterns as plotted in Fig. 7. The (103) magnetic reflection mainly arises from the magnetic ordering of the Nd sublattice. We note that the (103) magnetic reflection is overlapped with (112) nuclear Bragg peak as shown in Fig. 7. However, the nuclear structure factor of the (112) reflection is close to zero, and any structural transition at low temperature is excluded from our SXRPD data. Therefore, the sharp increase in the intensity of the (103) reflection between 2 and 10 K in the NPD data, as presented in the inset of Fig. 2(b), clearly suggests the magnetic ordering of the Nd-sublattice. Another indication of $Nd^{3+}$ ordering can be inferred from the substantial increase in intensities of magnetic reflections at higher Q (= $4\pi\sin\theta/\lambda$ ~ 2.51 Å$^{-1}$) where the magnetic form factor of $Ni^{2+}$ is low. Additional incommensurate superstructure reflections are also observed in the NPD data below 30 K. Potential origins of these incommensurate reflections are discussed in the next section.

We now discuss the commensurate magnetic structure of $Nd_2NiO_{4.11}$. The possible magnetic structures compatible with the $P4_2/ncm$ space group allowed by (100) propagation vector are obtained by applying representation theory using the program *BasIreps* included in *FullProf suite* program, the results are presented in the appendix. All the symmetry operations related to the $P4_2/ncm$ space group are listed in Table II while irreducible representations (IRs) of the space group with $\boldsymbol{k}$ = (100) are shown in Table III. After checking all of the possible magnetic modes obtained with representation analysis [*cf*. Table IV], the ferromagnetic component allowed for Ni or Nd sites is only valid in the one-dimensional $\Gamma_7$ magnetic mode, corresponding to the Shubnikov space group $P4_2/nc'm'$.

The neutron diffraction pattern collected at 2 K on DMC was used to determine the commensurate magnetic structure. The NPD data was refined using the $\Gamma_7$ magnetic model combining with the LTT structural model. The observed and calculated diffraction profiles are presented in Fig. 2(b) and the resulting magnetic structure is visualized in Fig. 6(b). Refined magnetic parameters are listed in table S-I of the Supplemental Material at [35]. In the $\Gamma_7$ magnetic mode, the spin direction in the basal plane coincides with the [110] tilt axis of the $NiO_6$ octahedron of the LTT phase. $Ni^{2+}$ and $Nd^{3+}$ moments are slightly canted away from the basal plane towards the *c*-axis to produce an effective ferromagnetic component. However, because of its small value, we were unable to refine the *c*-axis component of the magnetic moment with our NPD data. Its existence (~ 0.08 $\mu_B$/formula unit) was, however, unambiguously determined by means of isothermal magnetization measurements as discussed in the previous section. Refined ordered magnetic moments of Ni and Nd sites are 1.14(8) $\mu_B$ and 1.63(5) $\mu_B$ at 2 K, respectively. The refined magnetic moments of the $Ni^{2+}$ and $Nd^{3+}$ ions are respectively 28% and 50% smaller than the values reported for the stoichiometric compound (ordered magnetic moments of the $Ni^{2+}$ is 1.59 $\mu_B$ and of the $Nd^{3+}$ is 3.2 $\mu_B$ at 1.5 K) [25]. Please note that both, $Ni^{2+}$ and $Nd^{3+}$ ions are far from magnetic saturation even at 2 K in $Nd_2NiO_{4.11}$, as visible with the integrated intensity of (101) magnetic peak in Fig. 6(a). A possible explanation of reduced magnetic moment for the Ni-site could be deduced by considering the location of injected holes in the magnetic lattice due to oxygen doping. If the holes are Ni-site centered (that is, Ni $3d^7$), as in the case of site-centered stripes in $La_2NiO_{4+\delta}$ [30,58], the presence of $Ni^{3+}$ ions will reduce the saturated moment of the Ni-site. However, if the holes are bond centered (that is, Ni $3d^8\underline{L}$), i.e. occupying one of the equatorial oxygen site, as is the case of bond-centered stripes in $La_2NiO_{4+\delta}$ [30], the excess holes located at the oxygen sites cause a ferromagnetic coupling between the neighboring $Ni^{2+}$ spins, which is in competition with the normal antiferromagnetic superexchange interaction. The resulting frustration will also reduce the ordered moment of the Ni-site. In contrast, the reduction of the $Nd^{3+}$ magnetic moment is mainly due to changes in the crystalline electric field on the rare-earth site, as a consequence of double negatively charged $O_{int}$ ions in tetrahedral sites. These interstitial oxygen atoms strongly perturb the crystal-field levels of $Nd^{3+}$ ions. Even if a quantitative estimation of this effect is out of the scope of this work, important changes of the crystal-field splitting of the $J$ multiplet of the $Nd^{3+}$ are expected, which can affect the saturation value of the $Nd^{3+}$ magnetic moment.

In this context, it is important to point out the influence of the observed structural changes due to the oxygen over-stoichiometry on magnetic superexchange interaction parameters i.e. $J_{Ni-Ni}$; the intra-plane symmetric antiferromagnetic Ni-Ni superexchange interaction (see Fig. 6(b)) and $J'_{Ni-Ni}$; the interplane Ni-Ni superexchange interaction (see Fig. 6(b)). The $J_{Ni-Ni}$ interaction predominantly depends on two parameters: the superexchange angle ($\varphi$) and the superexchange distance ($d$) (see Fig. 6(c)). The tilting angle of $NiO_6$ octahedron decreases with oxygen doping, therefore, $\varphi$ changes from 158° to 163° and $d_{Ni\text{-}Oeq2}$ changes from 1.9668(6) Å to 1.9514(7) Å at 2 K. With these values, however, a relative increase in $J_{Ni\text{-}Ni}$ of 6% is expected on switching from $Nd_2NiO_{4.0}$ to $Nd_2NiO_{4.11}$. 3D magnetic order appears mainly because of the interplane magnetic interaction $J'_{Ni\text{-}Ni}$, which essentially depends on the Ni-$O_{ap}$ bond length and the distance between two nearest interplane apical oxygen atoms ($O_{ap}$-$O_{ap}$) through which superexchange occurs. The expansion of the rock salt (Nd-$O_{ap}$) layer and the change in the Ni-$O_{ap}$ bond length (only increased by 0.002 Å at 2 K) is not significant due to incorporation of excess oxygen atoms. Therefore, only a negligible change in $J'_{Ni\text{-}Ni}$ parameter is expected on switching from $Nd_2NiO_4$ to $Nd_2NiO_{4.11}$. It is evident that structural changes associated with oxygen doping cannot explain the observed reduction in 3D magnetic properties completely. Therefore, with our NPD study, we reveal that antiferromagnetic $J_{Ni\text{-}Ni}$ and $J'_{Ni\text{-}Ni}$ interactions are mainly influenced by the incorporated disorder in the magnetic lattice due to injected holes and not directly related to the interstitial oxygen atoms themselves like in the doped cuprates.

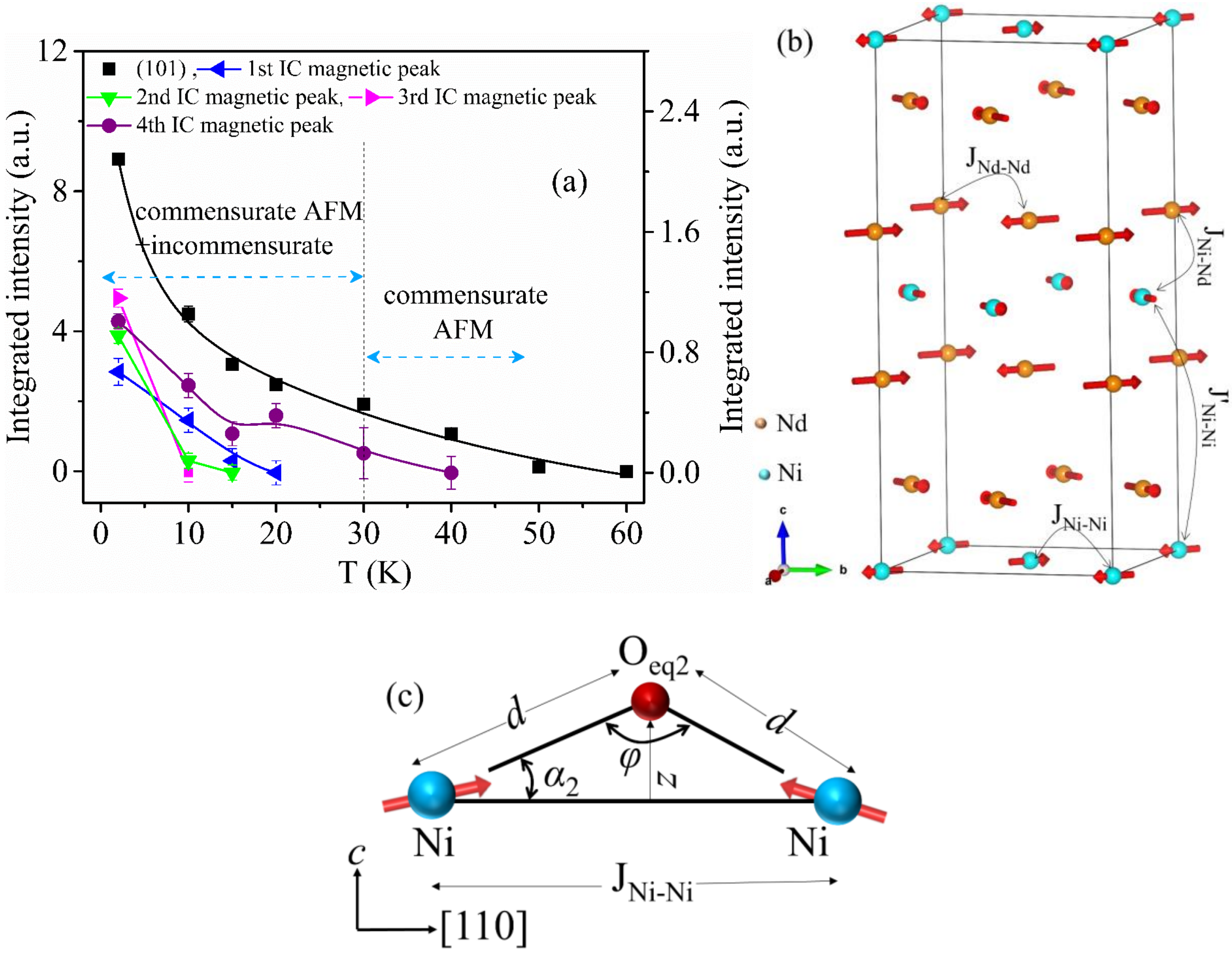

**Figure 6:** (a) Temperature evolution of integrated intensities of the (101) magnetic reflection (left side scale) and four intense incommensurate magnetic reflections (right side scale, IC1 to IC4) are plotted for $Nd_2NiO_{4.11}$ compound obtained with the NPD data. The vertical dashed line shows the temperature at which incommensurate reflections start to appear. (b) Commensurate antiferromagnetic structure of $Nd_2NiO_{4.11}$ compound in $P4_2/nc'm'$ Shubnikov space group. Only the Ni and Nd magnetic sites are shown. Ferromagnetic components of Ni and Nd moments along the *c*-axis are not shown for simplification. The in-plane antiferromagnetic superexchange $J_{Ni\text{-}Ni}$ occurs through Ni-$O_{eq}$-Ni bonds. The out of plane nearest neighbor Ni-Ni superexchange interaction $J'_{Ni\text{-}Ni}$ occurs through bridging Nd, $O_{ap}$ and $O_{int}$ atoms. The $J_{Ni\text{-}Nd}$ superexchange interaction goes through the apical site. (c) Breaking of the rotational symmetry regarding the Ni-Ni axis due to the displacement of $O_{eq2}$ atoms from the basal plane which allows the antisymmetric exchange interaction. Consequently, a weak

ferromagnetic component appears parallel to the *c*-axis in the LTT phase. The superexchange angle ($\varphi$) and distance ($d$) are also shown.

In addition to the commensurate magnetic reflections, additional weak incommensurate reflections are observed at $T \leq 30$ K. One potential origin of these incommensurate reflections could arise from a 2D/3D ordering of excess oxygen atoms, as previously reported for oxygen intercalated Ruddlesden-Popper type cuprates, nickelates and cobaltes [16,31,33,55,59]. However, the ordering temperature of excess oxygen atoms in these compounds, had usually been found to be close to 300 K. Therefore, the appearance of these incommensurate reflections in our data is likely not due to such oxygen ordering taking place at 30 K. A closer inspection reveals that they only have intensities in the low $q$ regime of the NPD data. Furthermore, these reflections are not present in our synchrotron XRPD data (see the inset of Fig. 7), strongly suggesting a magnetic origin. The positions of these incommensurate reflections could, however, not be related to wave vectors, associated to stripe magnetic order as reported for oxygen intercalated $La_2NiO_{4+\delta}$ compounds [32]. They are supposed to be related with the magnetic ordering of the $Ni^{2+}$, $Ni^{3+}$ and $Nd^{3+}$ spins. In Fig. 6(a) the integrated intensities of four intense incommensurate reflections as a function of temperature are outlined, as observed from the NPD data. The fact that the second and third incommensurate magnetic peaks appear below, whereas the first and fourth magnetic reflections appear above 10 K, is most probably related to $Nd^{3+}$ and $Ni^{2+}/Ni^{3+}$ ordering respectively. A quantitative analysis of the incommensurate magnetic phase is, however, beyond this powder diffraction work. A more dedicated single crystal neutron diffraction study combined with neutron polarization analysis is required to uncover the incommensurate phase detected in oxygen over-stoichiometric $Nd_2NiO_{4.11}$.

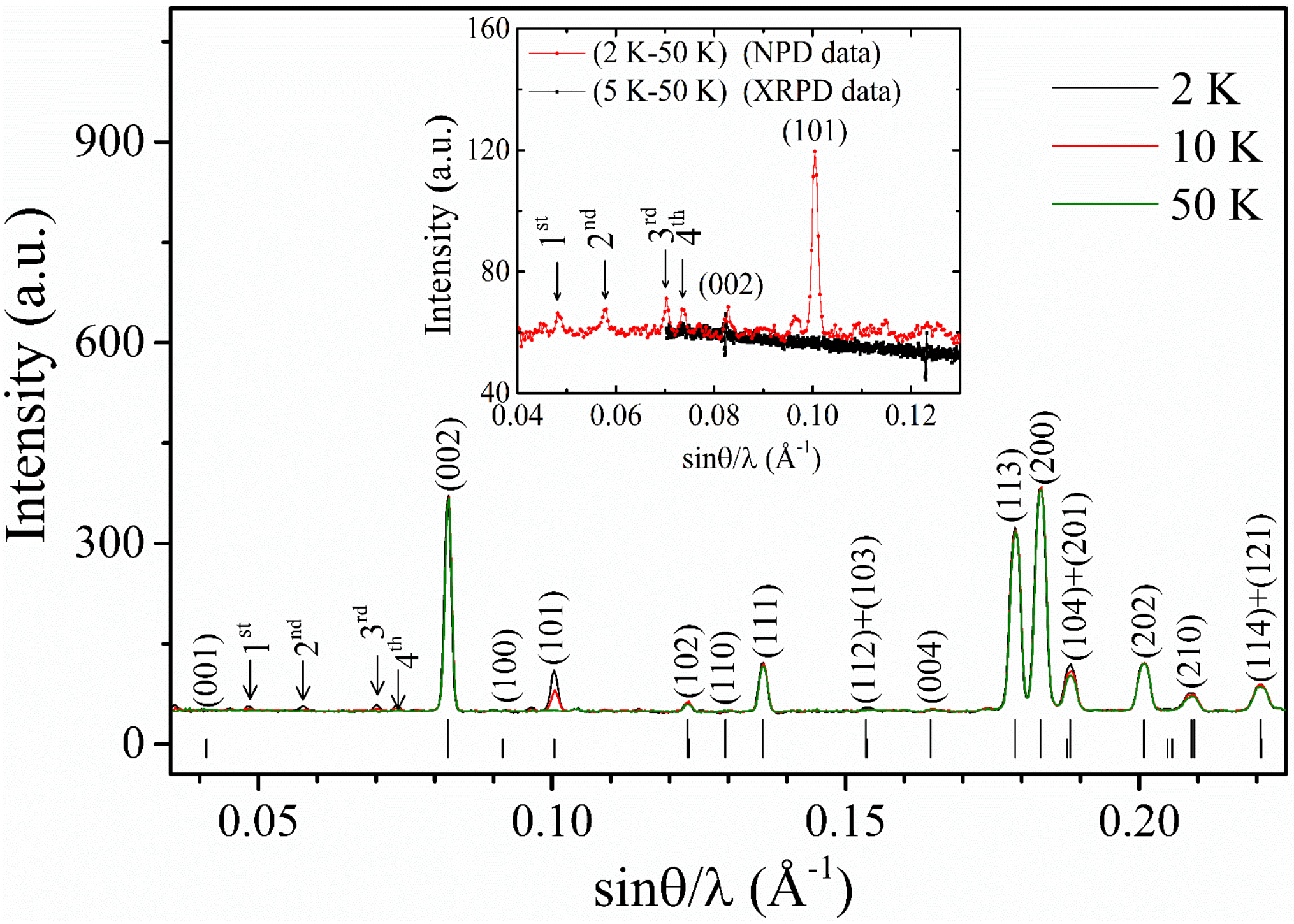

**Figure 7:** Temperature evolution of NPD patterns collected on DMC at SINQ, PSI ($\lambda$ = 2.4586(3) Å) indicating the appearance of magnetic order around 50 K. Only the low $q$ regime up to 0.25 Å$^{-1}$ is presented. Nuclear Bragg peaks are indicated by tick marks in the upper row while the lower row represents magnetic Bragg peaks with a (100) propagation vector. Indices of nuclear and commensurate magnetic reflections are given. Positions of four incommensurate reflections are shown by arrows. The inset shows the difference pattern obtained by subtracting 50 K data from 5 K data for XRPD (black circles) and subtracting 50 K data from 2 K data for NPD (red circles) emphasizing the magnetic origin of the incommensurate reflections.

## IV. CONCLUSIONS

The subtle modifications in the crystal structure and magnetic properties as a function of oxygen content ($\delta$) in Ruddlesden-Popper $K_2NiF_4$-type nickelates are not yet fully understood. In this context, we have addressed the influence of oxygen over-stoichiometry $\delta$ on the crystal structure and magnetic properties of $Nd_2NiO_{4.11}$ in the temperature range of 2-300 K. We revealed that for $Nd_2NiO_{4.11}$ the extra oxygen atoms selectively occupy the regular interstitial *4b* Wyckoff position with site symmetry $\bar{4}$ only. Thereby only one half out of four tetrahedral sites is statistically occupied to 11%. Occupied tetrahedra show a repulsive $O_{int}$-$O_{ap}$ interaction, leading to a symmetrical increase of the $O_{int}(O_{ap})_4$ tetrahedra, while the $O_{int}$-$O_{ap}$ distance reaches 2.76Å, instead of 2.15 Å when taking into account the average structure. Oxygen diffusion along the [110]-direction thus involves important local distortions of the respective tetrahedral sites, resembling to a successively in- and exhaling behaviour of the empty and occupied tetrahedra along the diffusion pathway. It is evident that such a mechanism is far from a rigid $O_{int}(O_{ap})_4$ configuration but that the associated structural changes of the tetrahedral sites during oxygen diffusion which considerably reduces any oxygen mobility in the $P4_2/ncm$ phase, compared to the disordered parent phase at high temperatures in F4/*mmm*, where all tetrahedra become dynamically activated and structurally equivalent. Moreover, strong anharmonic displacements of the apical oxygen atoms towards the nearest vacant interstitial sites are observed, induced by the presence of interstitial oxygen atoms. Such anharmonic behavior of the apical oxygens was interpreted to be an important prerequisite for a phonon assisted oxygen diffusion mechanism in $K_2NiF_4$ oxides close to room-temperature.

The oxygen over-stoichiometry, more precisely the incorporated holes (~ 1 hole per one unit cell in $Nd_2NiO_{4.11}$) strongly modify the 3D magnetic properties of $Nd_2NiO_{4.11}$ system. The Ni-sublattice orders antiferromagnetically at 53 K showing a weak ferromagnetic component along the *c*-axis, while Nd-sublatice orders only below 10 K. The magnetic structure at 2 K is, however, far more complex and consists of a commensurate phase with propagation vector (100) and an incommensurate phase. The incommensurate phase appears below 30 K and is different to what is previously associated with stripe or checkerboard type spin ordering. We assume that the incommensurate phase is likely to be related with the long-range ordering of the $Nd^{3+}$, $Ni^{2+}$ and $Ni^{3+}$ spins in the oxygen doped $Nd_2NiO_{4.11}$ phase and sensitively modulated by the charge transfer, i.e. the oxygen stoichiometry and ordering.

## V. ACKNOWLEDGEMENTS

This work is based on experiments performed at Paul Scherrer Institute, Villigen, Switzerland and at the Heinz Maier-Leibnitz Zentrum (MLZ), Garching, Germany. The authors acknowledge the beam times used at Material Sciences beamline X04SA at Swiss light source, on HRPT and DMC at PSI and on HEiDi at MLZ/FRM II (operated jointly by RWTH Aachen University and FZ Jülich within JARA collaboration) as well as laboratory equipment and support from LDM/PSI and the "Plateforme d'Analyse et de Caractérisation" of the ICG Montpellier. The authors gratefully acknowledge the financial support from the Swiss National Science Foundation (SNF) through grant 200021L_157131 and the French National Research Agency (ANR) through grant 14-CE36-0006-01 of the SECTOR project.

## APPENDIX: POSSIBLE MAGNETIC STRUCTURES BASED ON GROUP THEORY

The possible magnetic structures compatible with the crystal symmetry are obtained by applying the representation theory. The little group $G_k$ is a subset of symmetry elements within the paramagnetic space group $G_0$(P$4_2$/*ncm*), which leaves the propagation wave vector invariant under the unitary transformation matrix M. In the present case for ***k*** = (100), the little group contains all elements of $G_0$, which are listed in Table II. It is convenient to transform the representation of $G_k$ into irreducible representations (IRs) which are orthogonal to one another. The character table of $G_k$ is shown in table III. The magnetic representation at the Ni and Nd sites can be decomposed into a direct sum of irreducible representations as following:

$$\Gamma_{mag}(Ni) = 1\ \Gamma_1 + 2\ \Gamma_3 + 1\ \Gamma_5 + 2\ \Gamma_7 + 3\ \Gamma_{10}$$

$$\Gamma_{mag}(Nd) = 1\ \Gamma_1 + 2\ \Gamma_2 + 2\ \Gamma_3 + 1\ \Gamma_4 + 1\ \Gamma_5 + 2\ \Gamma_6 + 2\ \Gamma_7 + 1\ \Gamma_8 + 3\ \Gamma_9 + 3\ \Gamma_{10}$$

All IRs are one-dimensional, except $\Gamma_9$ and $\Gamma_{10}$, which are two-dimensional. The spin distribution of the *j*-th atom can be expressed as the Fourier transform of the linear combination of basis vectors, such that for a single propagation wave vector ***k***,

$$S_j = \sum_n C_n V_n e^{-ik.t} + c.c,$$

where the coefficients $C_n$ can, in general, be complex. There are four Ni (0 0 0) atoms and eight Nd (*x x z*) atoms within the primitive unit cell. They are $Ni_1$ (0 0 0), $Ni_2$ (1/2 1/2 0), $Ni_3$ (0 1/2 1/2), $Ni_4$ (1/2 0 1/2). Nd atoms are $Nd_1$ (*x*, *x*, *z*), $Nd_2$ (-*x*+1/2, -*x*+1/2, *z*), $Nd_3$ (-*x*, *x*+1/2, -*z*+1/2), $Nd_4$ (*x*+1/2, -*x*+1, -*z*+1/2), $Nd_5$ (*x*+1/2, *x*+1/2,-*z*), $Nd_6$ (-*x*, -*x*+1, -z), $Nd_7$ (*x*, -*x*+1/2, *z*+1/2) and $Nd_8$ (-*x*+1/2, *x*, *z*+1/2). The basis vectors belonging to each irreducible representation are presented in Table IV.

**Table II*:*** Symmetry operators of space group $P4_2/ncm$ showing explicitly the rotational part, IT notation as listed in the International Tables of Crystallography.

| Symbol | IT notation | Symbol | IT notation |
|---|---|---|---|
| $g_1$ | 1 | $g_9$ | -1 0,0,0 |
| $g_2$ | 2 1/4,1/4,*z* | $g_{10}$ | *n* (1/2,1/2,0) *x*,*y*,0 |
| $g_3$ | 2 (0,1/2,0) 0,*y*,1/4 | $g_{11}$ | *c x*,1/4,*z* |
| $g_4$ | 2 (1/2,0,0) *x*,1/2,1/4 | $g_{12}$ | *c* 1/4,y,*z* |
| $g_5$ | 2 (1/2,1/2,0) *x*,*x*,0 | $g_{13}$ | *m* x,-x+1/2,*z* |
| $g_6$ | 2 (-1/2,1/2,0) *x*,-*x*+1/2,0 | $g_{14}$ | *m x*,*x*,*z* |
| $g_7$ | 4- (0,0,1/2) 1/4,1/4,*z* | $g_{15}$ | -4- -1/4,1/4,*z*; -1/4,1/4,1 |
| $g_8$ | 4+ (0,0,1/2) 1/4,1/4,*z* | $g_{16}$ | -4+ 1/4,-1/4,*z*; 1/4,-1/4,1 |

**Table III:** Irreducible representations of the space group P4$_2$/*ncm* with ***k*** = (1 0 0). The final column gives the magnetic space group of each 1D IR in the Belov-Neronova-Smirnova notation.

| Γ | $g_1$ | $g_2$ | $g_3$ | $g_4$ | $g_5$ | $g_6$ | $g_7$ | $g_8$ | $g_9$ | $g_{10}$ | $g_{11}$ | $g_{12}$ | $g_{13}$ | $g_{14}$ | $g_{15}$ | $g_{16}$ | *M.S.G* |
|---|---|---|---|---|---|---|---|---|---|---|---|---|---|---|---|---|---|
| 1 | 1 | 1 | 1 | 1 | 1 | 1 | 1 | 1 | 1 | 1 | 1 | 1 | 1 | 1 | 1 | 1 | $P4_2/ncm$ |
| 2 | 1 | 1 | 1 | 1 | 1 | 1 | 1 | 1 | -1 | -1 | -1 | -1 | -1 | -1 | -1 | -1 | $P4_2/n'c'm'$ |
| 3 | 1 | 1 | 1 | 1 | -1 | -1 | -1 | -1 | 1 | 1 | 1 | 1 | -1 | -1 | -1 | -1 | $P4'_2/ncm'$ |
| 4 | 1 | 1 | 1 | 1 | -1 | -1 | -1 | -1 | -1 | -1 | -1 | -1 | 1 | 1 | 1 | 1 | $P4'_2/n'c'm$ |
| 5 | 1 | 1 | -1 | -1 | 1 | 1 | -1 | -1 | 1 | 1 | -1 | -1 | 1 | 1 | -1 | -1 | $P4'_2/nc'm$ |
| 6 | 1 | 1 | -1 | -1 | 1 | 1 | -1 | -1 | -1 | -1 | 1 | 1 | -1 | -1 | 1 | 1 | $P4'_2/n'cm'$ |
| 7 | 1 | 1 | -1 | -1 | -1 | -1 | 1 | 1 | 1 | 1 | -1 | -1 | -1 | -1 | 1 | 1 | $P4_2/nc'm'$ |
| 8 | 1 | 1 | -1 | -1 | -1 | -1 | 1 | 1 | -1 | -1 | 1 | 1 | 1 | 1 | -1 | -1 | $P4_2/n'cm$ |
| 9 | $\begin{matrix}1&0\\0&1\end{matrix}$ | $\begin{matrix}-1&0\\0&-1\end{matrix}$ | $\begin{matrix}1&0\\0&-1\end{matrix}$ | $\begin{matrix}-1&0\\0&1\end{matrix}$ | $\begin{matrix}0&1\\1&0\end{matrix}$ | $\begin{matrix}0&-1\\-1&0\end{matrix}$ | $\begin{matrix}0&-1\\1&0\end{matrix}$ | $\begin{matrix}0&1\\-1&0\end{matrix}$ | $\begin{matrix}-1&0\\0&-1\end{matrix}$ | $\begin{matrix}1&0\\0&1\end{matrix}$ | $\begin{matrix}-1&0\\0&1\end{matrix}$ | $\begin{matrix}1&0\\0&-1\end{matrix}$ | $\begin{matrix}0&-1\\-1&0\end{matrix}$ | $\begin{matrix}0&1\\1&0\end{matrix}$ | $\begin{matrix}0&1\\-1&0\end{matrix}$ | $\begin{matrix}0&-1\\1&0\end{matrix}$ | --- |
| 10 | $\begin{matrix}1&0\\0&1\end{matrix}$ | $\begin{matrix}-1&0\\0&-1\end{matrix}$ | $\begin{matrix}1&0\\0&-1\end{matrix}$ | $\begin{matrix}-1&0\\0&1\end{matrix}$ | $\begin{matrix}0&1\\1&0\end{matrix}$ | $\begin{matrix}0&-1\\-1&0\end{matrix}$ | $\begin{matrix}0&-1\\1&0\end{matrix}$ | $\begin{matrix}0&1\\-1&0\end{matrix}$ | $\begin{matrix}1&0\\0&1\end{matrix}$ | $\begin{matrix}-1&0\\0&-1\end{matrix}$ | $\begin{matrix}1&0\\0&-1\end{matrix}$ | $\begin{matrix}-1&0\\0&1\end{matrix}$ | $\begin{matrix}0&1\\1&0\end{matrix}$ | $\begin{matrix}0&-1\\-1&0\end{matrix}$ | $\begin{matrix}0&-1\\1&0\end{matrix}$ | $\begin{matrix}0&1\\-1&0\end{matrix}$ | --- |

**Table IV:** Basis vectors corresponding of the Ni and Nd atoms. The notation for the direction is [$\boldsymbol{e_x}$ $\boldsymbol{e_y}$ $\boldsymbol{e_z}$]; $\boldsymbol{e_z}$ is parallel to the *c*-axis axis and $\boldsymbol{e_x}$ and $\boldsymbol{e_y}$ are in the basal plane parallel to the *a*- and *b*-axis, respectively.

| IRs | Basis vectors | $Ni_1$, $Nd_1$ | $Ni_2$, $Nd_2$ | $Ni_3$, $Nd_3$ | $Ni_4$, $Nd_4$ | $Nd_5$ | $Nd_6$ | $Nd_7$ | $Nd_8$ |
|---|---|---|---|---|---|---|---|---|---|
| $\Gamma_1$ | $V_1^1$ | [1 $\bar{1}$ 0] | [$\bar{1}$ 1 0] | [$\bar{1}$ $\bar{1}$ 0] | [1 1 0] | [$\bar{1}$ 1 0] | [1 $\bar{1}$ 0] | [$\bar{1}$ $\bar{1}$ 0] | [1 1 0] |
| $\Gamma_2$ | $V_2^1$ | [1 1 0] | [$\bar{1}$ $\bar{1}$ 0] | [$\bar{1}$ 1 0] | [1 $\bar{1}$ 0] | [1 1 0] | [$\bar{1}$ $\bar{1}$ 0] | [1 $\bar{1}$ 0] | [$\bar{1}$ 1 0] |
| | $V_2^2$ | [0 0 1] | [0 0 1] | [0 0 $\bar{1}$] | [0 0 $\bar{1}$] | [0 0 $\bar{1}$] | [0 0 $\bar{1}$] | [0 0 1] | [0 0 1] |
| $\Gamma_3$ | $V_3^1$ | [1 1 0] | [$\bar{1}$ $\bar{1}$ 0] | [$\bar{1}$ 1 0] | [1 $\bar{1}$ 0] | [$\bar{1}$ $\bar{1}$ 0] | [1 1 0] | [$\bar{1}$ 1 0] | [1 $\bar{1}$ 0] |
| | $V_3^2$ | [0 0 1] | [0 0 1] | [0 0 $\bar{1}$] | [0 0 $\bar{1}$] | [0 0 1] | [0 0 1] | [0 0 $\bar{1}$] | [0 0 $\bar{1}$] |
| $\Gamma_4$ | $V_4^1$ | [1 $\bar{1}$ 0] | [$\bar{1}$ 1 0] | [$\bar{1}$ $\bar{1}$ 0] | [1 1 0] | [1 $\bar{1}$ 0] | [$\bar{1}$ 1 0] | [1 1 0] | [$\bar{1}$ $\bar{1}$ 0] |
| $\Gamma_5$ | $V_5^1$ | [1 $\bar{1}$ 0] | [$\bar{1}$ 1 0] | [1 1 0] | [$\bar{1}$ $\bar{1}$ 0] | [$\bar{1}$ 1 0] | [1 $\bar{1}$ 0] | [1 1 0] | [$\bar{1}$ $\bar{1}$ 0] |
| $\Gamma_6$ | $V_6^1$ | [1 1 0] | [$\bar{1}$ $\bar{1}$ 0] | [1 $\bar{1}$ 0] | [$\bar{1}$ 1 0] | [1 1 0] | [$\bar{1}$ $\bar{1}$ 0] | [$\bar{1}$ 1 0] | [1 $\bar{1}$ 0] |
| | $V_6^2$ | [0 0 $\bar{1}$] | [0 0 $\bar{1}$] | [0 0 $\bar{1}$] | [0 0 $\bar{1}$] | [0 0 $\bar{1}$] | [0 0 $\bar{1}$] | [0 0 $\bar{1}$] | [0 0 $\bar{1}$] |
| $\Gamma_7$ | $V_7^1$ | [1 1 0] | [$\bar{1}$ $\bar{1}$ 0] | [1 $\bar{1}$ 0] | [$\bar{1}$ 1 0] | [1 1 0] | [$\bar{1}$ $\bar{1}$ 0] | [$\bar{1}$ 1 0] | [1 $\bar{1}$ 0] |
| | $V_7^2$ | [0 0 1] | [0 0 1] | [0 0 1] | [0 0 1] | [0 0 1] | [0 0 1] | [0 0 1] | [0 0 1] |
| $\Gamma_8$ | $V_8^1$ | [1 $\bar{1}$ 0] | [$\bar{1}$ 1 0] | [1 1 0] | [$\bar{1}$ $\bar{1}$ 0] | [1 $\bar{1}$ 0] | [$\bar{1}$ 1 0] | [$\bar{1}$ $\bar{1}$ 0] | [1 1 0] |
| $\Gamma_9$ | $V_9^1$ | [1 0 0] | [1 0 0] | $[\bar{1}\ 0\ 0]$ | $[\bar{1}\ 0\ 0]$ | $[\bar{1}\ 0\ 0]$ | $[\bar{1}\ 0\ 0]$ | [1 0 0] | [1 0 0] |
| | $V_9^2$ | [0 1 0] | [0 1 0] | [0 1 0] | [0 1 0] | $[0\ \bar{1}\ 0]$ | $[0\ \bar{1}\ 0]$ | $[0\ \bar{1}\ 0]$ | $[0\ \bar{1}\ 0]$ |
| | $V_9^3$ | [0 0 1] | $[0\ 0\ \bar{1}]$ | $[0\ 0\ \bar{1}]$ | [0 0 1] | [0 0 1] | $[0\ 0\ \bar{1}]$ | [0 0 1] | $[0\ 0\ \bar{1}]$ |
| | $V_9^4$ | $[0\ \bar{1}\ 0]$ | $[0\ \bar{1}\ 0]$ | [0 1 0] | [0 1 0] | [0 1 0] | [0 1 0] | $[0\ \bar{1}\ 0]$ | $[0\ \bar{1}\ 0]$ |
| | $V_9^5$ | $[\bar{1}\ 0\ 0]$ | $[\bar{1}\ 0\ 0]$ | $[\bar{1}\ 0\ 0]$ | $[\bar{1}\ 0\ 0]$ | [1 0 0] | [1 0 0] | [1 0 0] | [1 0 0] |
| | $V_9^6$ | $[0\ 0\ \bar{1}]$ | [0 0 1] | $[0\ 0\ \bar{1}]$ | [0 0 1] | $[0\ 0\ \bar{1}]$ | [0 0 1] | [0 0 1] | $[0\ 0\ \bar{1}]$ |
| $\Gamma_{10}$ | $V_{10}^1$ | [1 0 0] | [1 0 0] | $[\bar{1}\ 0\ 0]$ | $[\bar{1}\ 0\ 0]$ | [1 0 0] | [1 0 0] | $[\bar{1}\ 0\ 0]$ | $[\bar{1}\ 0\ 0]$ |
| | $V_{10}^2$ | [0 1 0] | [0 1 0] | [0 1 0] | [0 1 0] | [0 1 0] | [0 1 0] | [0 1 0] | [0 1 0] |
| | $V_{10}^3$ | [0 0 1] | $[0\ 0\ \bar{1}]$ | $[0\ 0\ \bar{1}]$ | [0 0 1] | $[0\ 0\ \bar{1}]$ | [0 0 1] | $[0\ 0\ \bar{1}]$ | [0 0 1] |
| | $V_{10}^4$ | [0 1 0] | [0 1 0] | $[0\ \bar{1}\ 0]$ | $[0\ \bar{1}\ 0]$ | [0 1 0] | [0 1 0] | $[0\ \bar{1}\ 0]$ | $[0\ \bar{1}\ 0]$ |
| | $V_{10}^5$ | [1 0 0] | [1 0 0] | [1 0 0] | [1 0 0] | [1 0 0] | [1 0 0] | [1 0 0] | [1 0 0] |
| | $V_{10}^6$ | [0 0 1] | $[0\ 0\ \bar{1}]$ | [0 0 1] | $[0\ 0\ \bar{1}]$ | $[0\ 0\ \bar{1}]$ | [0 0 1] | [0 0 1] | $[0\ 0\ \bar{1}]$ |

[1] J. Bassat, M. Burriel, O. Wahyudi, R. Castaing, M. Ceretti, P. Veber, I. Weill, A. Villesuzanne, J. Grenier, W. Paulus, and J. A. Kilner, J. Phys. Chem. C **117**, 26466 (2013).

[2] D. Lee and H. N. Lee, Materials (Basel). **10**, 1 (2017).

[3] J.-M. Bassat, M. Burriel, M. Ceretti, P. Veber, J.-C. Grenier, W. Paulus, and J. A. Kilner, ECS Trans. **57**, 1753 (2013).

[4] J. A. Kilner and M. Burriel, Annu. Rev. Mater. Res. **44**, 365 (2014).

[5] J. Wan, J. B. Goodenough, and J. H. Zhu, Solid State Ionics **178**, 281 (2007).

[6] C. Ferchaud, J. C. Grenier, Y. Zhang-Steenwinkel, M. M. A. Van Tuel, F. P. F. Van Berkel, and J. M. Bassat, J. Power Sources **196**, 1872 (2011).

[7] X. Tong, F. Zhou, S. Yang, S. Zhong, M. Wei, and Y. Liu, Ceram. Int. **43**, 10927 (2017).

[8] M. A. Laguna-Bercero, H. Monzón, A. Larrea, and V. M. Orera, J. Mater. Chem. A **4**, 1446 (2016).

[9] A. V. Kovalevsky, V. V. Kharton, A. A. Yaremchenko, Y. V. Pivak, E. V. Tsipis, S. O. Yakovlev, A. A. Markov, E. N. Naumovich, and J. R. Frade, J. Electroceramics **18**, 205 (2007).

[10] M. Ceretti, O. Wahyudi, G. André, M. Meven, A. Villesuzanne, and W. Paulus, Inorg. Chem. **57**, 4657 (2018).

[11] S. Bhavaraju, J. F. DiCarlo, D. P. Scarfe, I. Yazdi, and A. J. Jacobson, Chem. Mater. **6**, 2172 (1994).

[12] A. Perrichon, A. Piovano, M. Boehm, M. Zbiri, M. Johnson, H. Schober, M. Ceretti, and W. Paulus, J. Phys. Chem. C **119**, 1557 (2015).

[13] A. Villesuzanne, W. Paulus, A. Cousson, S. Hosoya, L. Le Dréau, O. Hernandez, C. Prestipino, M. Ikbel Houchati, and J. Schefer, J. Solid State Electrochem. **15**, 357 (2011).

[14] D. Parfitt, A. Chroneos, J. A. Kilner, and R. W. Grimes, Phys. Chem. Chem. Phys. **12**, 6834 (2010).

[15] W. Paulus, H. Schober, S. Eibl, M. Johnson, T. Berthier, O. Hernandez, M. Ceretti, M. Plazanet, K. Conder, and C. Lamberti, J. Am. Chem. Soc. **130**, 16080 (2008).

[16] M. Ceretti, O. Wahyudi, A. Cousson, A. Villesuzanne, M. Meven, B. Pedersen, J. M. Bassat, and W. Paulus, J. Mater. Chem. A **3**, 21140 (2015).

[17] A. Piovano, A. Perrichon, M. Boehm, M. R. Johnson, and W. Paulus, Phys. Chem. Chem. Phys. **18**, 17398 (2016).

[18] K. Ishikawa, K. Metoki, and H. Miyamoto, J. Solid State Chem. **182**, 2096 (2009).

[19] H. Ishikawa, Y. Toyosumi, and K. Ishikawa, J. Alloys Compd. **408–412**, 1196 (2006).

[20] M. T. Fernández-Díaz, J. L. Martínez, and J. Rodríguez-Carvajal, Solid State Ionics **63–65**, 902 (1993).

[21] M. Castro and R. Burriel, Thermochim. Acta **269–270**, 523 (1995).

[22] K. Ishikawa, Solid State Ionics **262**, 682 (2014).

[23] A. Flura, S. Dru, C. Nicollet, V. Vibhu, S. Fourcade, E. Lebraud, A. Rougier, J.-M. Bassat, and

J.-C. Grenier, J. Solid State Chem. **228**, 189 (2015).

[24] M. Zaghrioui, F. Giovannelli, N. P. D. Brouri, and I. Laffez, J. Solid State Chem. **177**, 3351 (2004).

[25] J. Rodrguez Carvajal, M. T. Fernandez Daz, J. L. Martnez, F. Ferndez, and R. Saez Puche, Epl **11**, 261 (1990).

[26] L. C. Otero-Diaz, A. R. Landa, F. Fernandez, R. Saez-Puche, R. Withers, and B. G. Hyde, J. Solid State Chem. **97**, 443 (1992).

[27] D. J. Buttrey and J. M. Honig, J. Solid State Chem. **72**, 38 (1988).

[28] K. Nakajima, Y. Endoh, S. Hosoya, J. Wada, D. Welz, H. M. Mayer, H. A. Graf, and M. Steiner, J. Phys. Soc. Japan **66**, 809 (1997).

[29] K. Yamada, T. Omata, K. Nakajima, Y. Endoh, and S. Hosoya, Phys. C Supercond. Its Appl. **221**, 355 (1994).

[30] P. Wochner, J. M. Tranquada, D. J. Buttrey, and V. Sachan, Phys. Rev. B **57**, 1066 (1998).

[31] Z. Hiroi, T. Obata, M. Takano, Y. Bando, Y. Takeda, and O. Yamamoto, Phys. Rev. B **41**, 11665 (1990).

[32] J. M. Tranquada, D. J. Buttrey, V. Sachan, and J. E. Lorenzo, Phys. Rev. Lett. **73**, 1003 (1994).

[33] J. M. Tranquada, Y. Kong, J. E. Lorenzo, D. J. Buttrey, D. E. Rice, and V. Sachan, Phys. Rev. B **50**, 6340 (1994).

[34] O. Wahyudi, M. Ceretti, I. Weill, A. Cousson, F. Weill, M. Meven, M. Guerre, A. Villesuzanne, J.-M. Bassat, and W. Paulus, CrystEngComm **17**, 6278 (2015).

[35] See the Supplemental Material for the analysis of the overall oxygen content in the powder sample by thermogravimetry analysis; synchrotron x-ray powder diffraction data collected at 5 K and 300 K; powder neutron diffraction patterns collected at 2 K and 300 K; and Rietveld refinement results obtained from powder neutron diffraction data.

[36] P. R. Willmott, D. Meister, S. J. Leake, M. Lange, A. Bergamaschi, M. Böge, M. Calvi, C. Cancellieri, N. Casati, A. Cervellino, Q. Chen, C. David, U. Flechsig, F. Gozzo, B. Henrich, S. Jäggi-Spielmann, B. Jakob, I. Kalichava, P. Karvinen, J. Krempasky, A. Lüdeke, R. Lüscher, S. Maag, C. Quitmann, M. L. Reinle-Schmitt, T. Schmidt, B. Schmitt, A. Streun, I. Vartiainen, M. Vitins, X. Wang, and R. Wullschleger, J. Synchrotron Radiat. **20**, 667 (2013).

[37] P. Fischer, G. Frey, M. Koch, M. Konnecke, V. Pomjakushin, J. Schefer, R. Thut, N. Schlumpf, R. Burge, U. Greuter, S. Bondt, and E. Berruyer, Phys. B Condens. Matter **276–278**, 146 (2000).

[38] J. Schefer, P. Fischer, H. Heer, A. Isacson, M. Koch, and R. Thut, Nucl. Instruments Methods Phys. Res. Sect. A Accel. Spectrometers, Detect. Assoc. Equip. **288**, 477 (1990).

[39] Cold Neutron Powder Diffractometer DMC. Available at: https://www.psi.ch/sinq/dmc/

[40] J. Rodríguez-Carvajal, Phys. B Phys. Condens. Matter **192**, 55 (1993).

[41] M. Meven and A. Sazonov, J. Large-Scale Res. Facil. JLSRF **1**, A7 (2015).

[42] V. Petrícek, M. Dušek, and L. Palatinus, Zeitschrift Fur Krist. **229**, 345 (2014).

[43] M. Sakata and M. Sato, Acta Crystallogr. Sect. A **46**, 263 (1990).

[44] D. M. Collins, Nature **298**, 49 (1982).

[45] K. Momma, T. Ikeda, A. A. Belik, and F. Izumi, Powder Diffr. **28**, 184 (2013).

[46] J. Nocedal, Deforestation Clim. Chang. **35**, 39 (2010).

[47] K. Momma and F. Izumi, J. Appl. Crystallogr. **41**, 653 (2008).

[48] K. Momma and F. Izumi, J. Appl. Crystallogr. **44**, 1272 (2011).

[49] J. D. Sullivan, D. J. Buttrey, D. E. Cox, and J. Hriljac, J. Solid State Chem. **94**, 337 (1991).

[50] T. Kiyama, K. Yoshimura, K. Kosuge, Y. Ikeda, and Y. Bando, Phys. Rev. B - Condens. Matter Mater. Phys. **54**, R756 (1996).

[51] P. G. Freeman, N. B. Christensen, D. Prabhakaran, and A. T. Boothroyd, J. Phys. Conf. Ser. **200**, 012037 (2010).

[52] R. D. Shannon, Acta Cryst. A **32**, 751 (1976).

[53] A. Chroneos, D. Parfitt, J. A. Kilner, and R. W. Grimes, J. Mater. Chem. **20**, 266 (2010).

[54] W. Paulus, A. Cousson, G. Dhalenne, J. Berthon, A. Revcolevschi, S. Hosoya, W. Treutmann, G. Heger, and R. Le Toquin, Solid State Sci. **4**, 565 (2002).

[55] L. Le Dréau, C. Prestipino, O. Hernandez, J. Schefer, G. Vaughan, S. Paofai, J. M. Perez-Mato, S. Hosoya, and W. Paulus, Inorg. Chem. **51**, 9789 (2012).

[56] X. Batlle, X. Obradors, and B. Martnez, Phys. Rev. B **45**, 2830 (1992).

[57] X. Obradors, X. Batlle, J. Rodrguez-Carvajal, J. L. Martnez, M. Vallet, J. González-Calbet, and J. Alonso, Phys. Rev. B **43**, 10451 (1991).

[58] M. Uchida, Y. Yamasaki, Y. Kaneko, K. Ishizaka, J. Okamoto, H. Nakao, Y. Murakami, and Y. Tokura, Phys. Rev. B - Condens. Matter Mater. Phys. **86**, (2012).

[59] B. O. Wells, R. J. Birgeneau, F. C. Chou, Y. Endoh, D. C. Johnston, M. A. Kastner, Y. S. Lee, G. Shirane, J. M. Tranquada, and K. Yamada, Zeitschrift Fur Phys. B-Condensed Matter **100**, 535 (1996).

Electronic Supplemental Material

# Structural disorder and magnetic correlations driven by oxygen doping in $Nd_2NiO_{4+\delta}$ ($\delta$ ~ 0.11)

Sumit Ranjan Maity,[1,2] Monica Ceretti,[3] Lukas Keller,[1] Jürg Schefer,[1] Tian Shang,[4,5,6] Ekaterina Pomjakushina,[4] Martin Meven,[7] Denis Sheptyakov,[1] Antonio Cervellino,[5] Werner Paulus[3]

[1]*Laboratory for Neutron Scattering and Imaging, Paul Scherrer Institut, Villigen CH-5232, Switzerland.*

[2]*University of Geneva, Department of Quantum Matter Physics (DQMP) 24, Quai Ernest Ansermet CH-1211 Genève 4, Switzerland*

[3]*Institut Charles Gerhardt, UMR 5253, CNRS-University Montpellier, 34095 Montpellier, France*

[4]*Laboratory for Multiscale Materials Experiments, Paul Scherrer Institut, Villigen CH-5232, Switzerland*

[5]*Swiss Light Source, Paul Scherrer Institut, Villigen CH-5232, Switzerland*

[6]*Institute of Condensed Matter Physics, École polytechnique fédérale de Lausanne (EPFL), Lausanne CH-1015, Switzerland*

[7]*Institute of Crystallography, RWTH Aachen University and Jülich Centre for Neutron Science (JCNS), Forschungszentrum Jülich GmbH at Heinz Maier-Leibnitz Zentrum (MLZ), 85747 Garching, Germany*

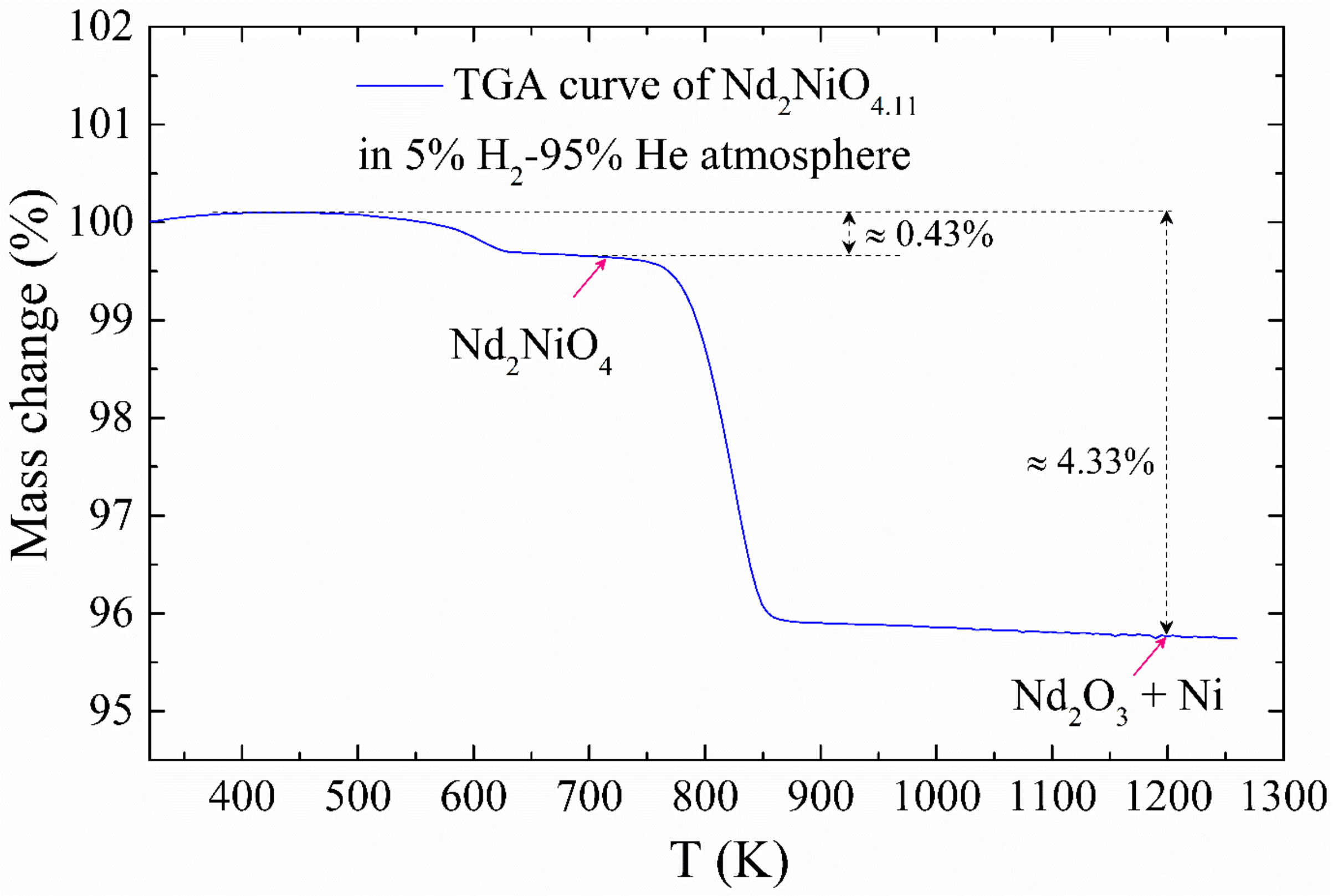

**Figure S-1:** TGA curve of $Nd_2NiO_{4.11}$ compound measured in 5%-$H_2$/95%-He gas atmosphere in the temperature range of 300-1250 K.

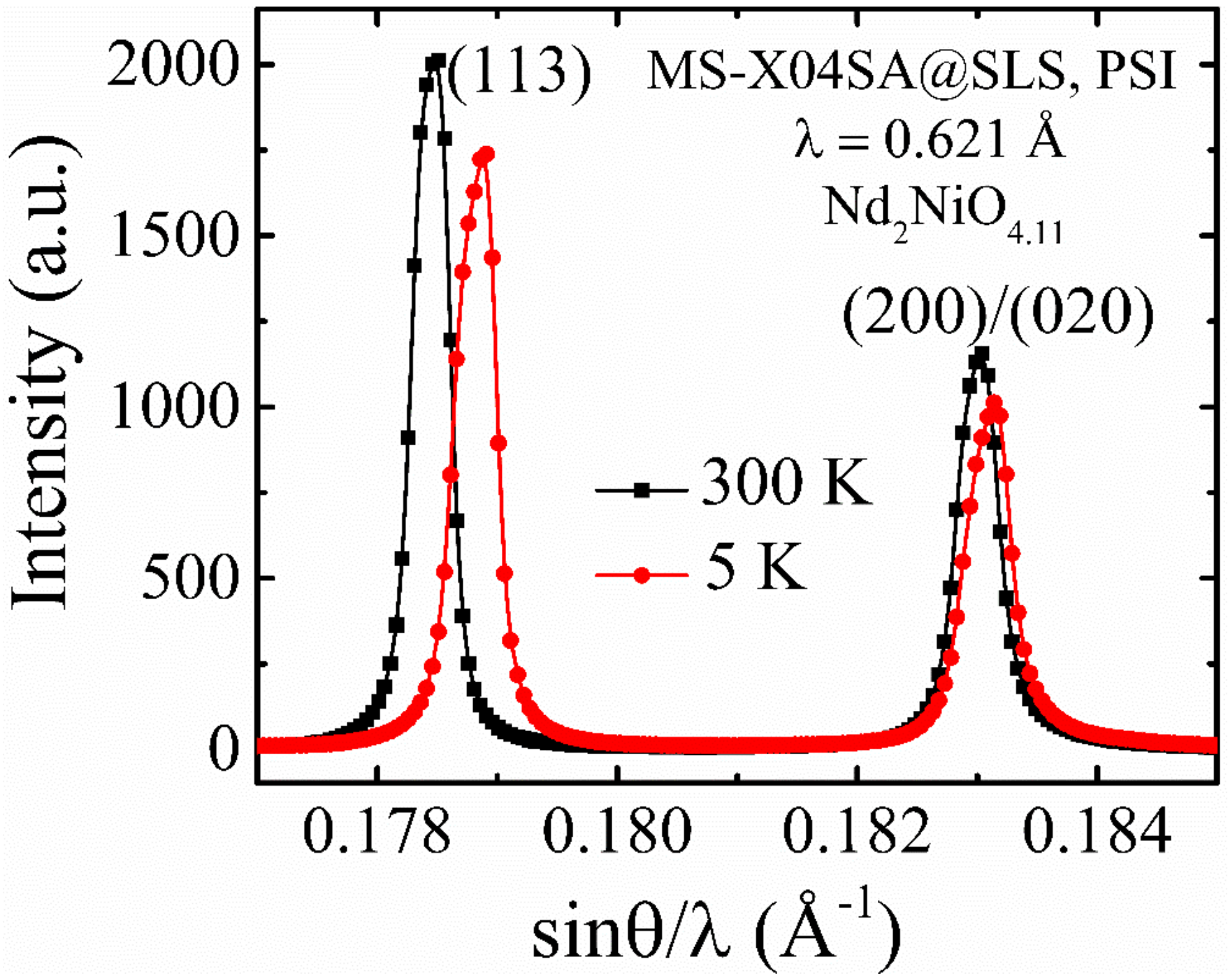

**Figure S-2:** A magnified portion of the high-resolution synchrotron X-ray diffraction data collected at 300 K and 5 K at the Material Sciences beamline X04SA of SLS, PSI. No splitting of (200) Bragg reflection indicates the absence orthorhombic distortion in the crystal structure of $Nd_2NiO_{4.11}$ down to 5 K.

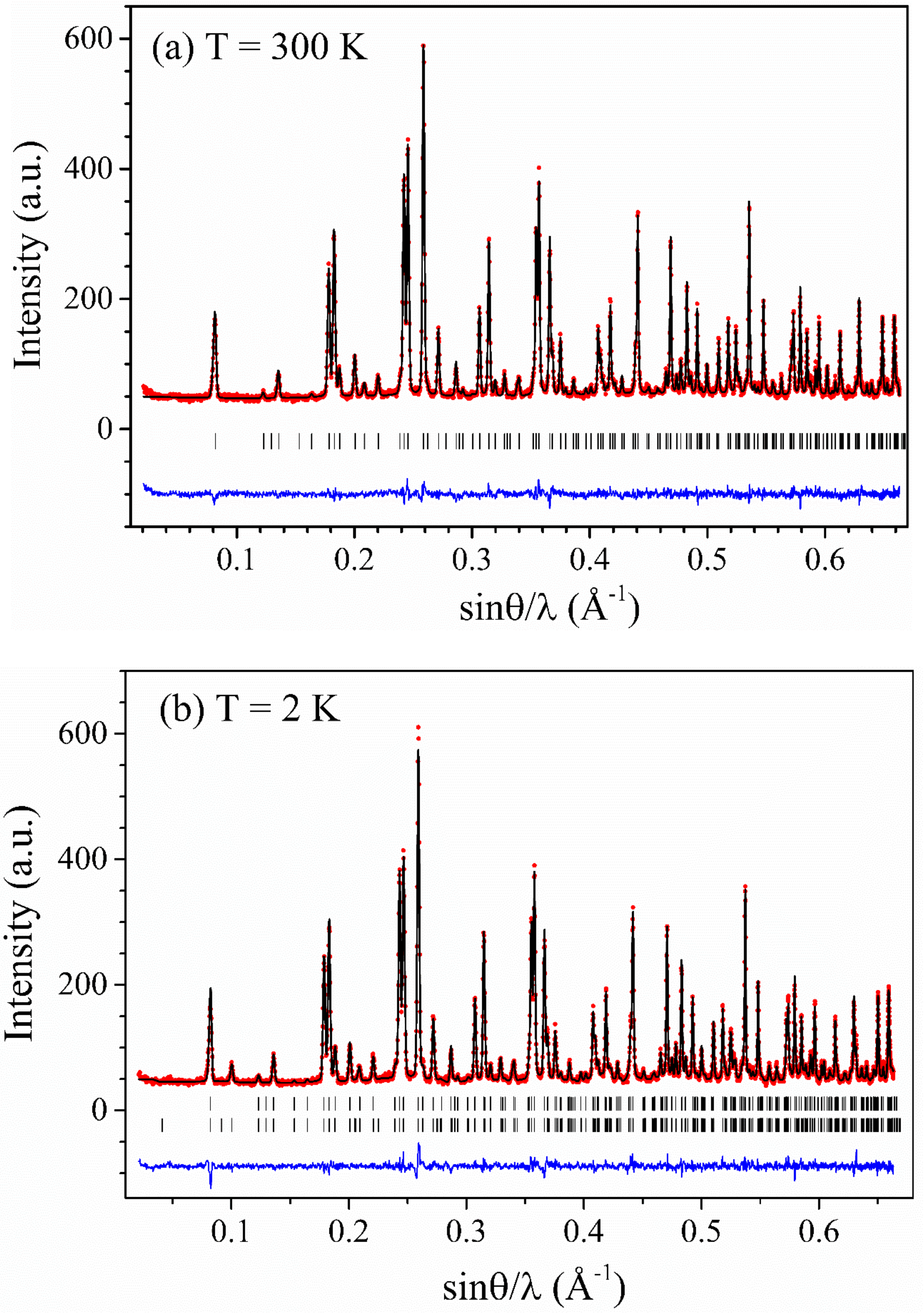

**Figure S-3:** Observed (red circles), calculated (black line) and difference (blue continuous line) patterns resulting from Rietveld analysis of neutron powder diffraction data for the $Nd_2NiO_{4.11}$ compound collected on HRPT at SINQ, PSI ($\lambda$ = 1.4940(2) Å) at (a) T = 300 K and (b) T = 2 K. All powder data are refined with Rietveld method. Nuclear Bragg reflections are marked with ticks in the upper row. Lower row of ticks in panel (b) gives the position of commensurate magnetic peaks with the propagation vector (100). The results are summarized in Table I.

**TABLE S-I:** Refinement results of polycrystalline $Nd_2NiO_{4.11}$ compound in P*4₂/ncm* space group using Rietveld method. Data collected on HRPT at SINQ, PSI, λ = 1.4940(2) Å up to $\sin\theta_{max}/\lambda$ = 0.66 $Å^{-1}$), Thermal displacement parameters $U_{ij}$ are given in ($Å^2$).

| Temperature | | 300 K | 2 K |
|---|---|---|---|
| *a* = *b* | (Å) | 5.46200(2) | 5.45868(1) |
| *c* | (Å) | 12.20517(9) | 12.15376(8) |
| Ni (0 0 0) | Occ. | 1 | 1 |
| | $U_{iso}$ | 0.0072(3) | 0.0039(2) |
| | $\mu_x = \mu_y$ ($\mu_B$) | --- | 0.81(5) |
| | $\mu_{total}$ ($\mu_B$) | --- | 1.14(7) |
| Nd (*x x z*) | Occ. | 2 | 2 |
| | *x* | 0.9889(3) | 0.98731(18) |
| | *z* | 0.36104(12) | 0.36104(9) |
| | $U_{iso}$ | 0.0096(4) | 0.0052(2 |
| | $\mu_x = \mu_y$ ($\mu_B$) | --- | 1.15(3) |
| | $\mu_{total}$ ($\mu_B$) | --- | 1.63(5) |
| $O_{ap}$ (*x x z*) | Occ. | 2.006(7) | 2.023(6) |
| | *x* | 0.0458(4) | 0.0473(3) |
| | *z* | 0.1783(3) | 0.1779(2) |
| | $U_{11} = U_{22}$ | 0.0382(14) | 0.0335(10) |
| | $U_{33}$ | 0.0085(11) | 0.0133(8) |
| | $U_{12}$ | -0.0023(14) | -0.0024(10) |
| | $U_{13} = U_{23}$ | 0.0009(10) | 0.0010(8) |
| $O_{eq1}$ (¾ ¼ 0) | Occ. | 1.01(3) | 0.966(20) |
| | $U_{11} = U_{22}$ | 0.0077(16) | 0.0100(10) |
| | $U_{33}$ | 0.041(4) | 0.0134(14) |
| | $U_{12}$ | 0.002(2) | -0.0007(15) |
| $O_{eq2}$ (¼ ¼ *z*) | Occ. | 0.97(3) | 1.021(20) |
| | *z* | 0.9795(4) | 0.9766(3) |
| | $U_{11} = U_{22}$ | 0.0095(16) | 0.0085(10) |
| | $U_{33}$ | 0.0119(19) | 0.0108(14) |
| | $U_{12}$ | -0.001(2) | 0.0006(15) |
| $O_{int}$ (¾ ¼ ¼) | Occ. | 0.11 | 0.11 |
| | $U_{iso}$ | 0.026(6) | 0.037(7) |
| $R_p$ (%) | | 15.5 | 12 |
| $wR_p$ (%) | | 14.2 | 11.2 |
| $R_{Bragg}$ (%) | | 2.86 | 3.45 |
| $R_{Magnetic}$ (%) | | --- | 7.42 |